\documentclass[lettersize,journal]{IEEEtran}
\usepackage{amsmath, amsfonts}
\usepackage{algorithmic}
\usepackage{array}
\usepackage[caption=false,font=normalsize,labelfont=sf,textfont=sf]{subfig}
\usepackage{textcomp}
\usepackage{stfloats}
\usepackage{url}
\usepackage{verbatim}
\usepackage{graphicx}

\usepackage{supertabular,booktabs}
\usepackage{multirow}
\usepackage{multicol}
\usepackage[figuresright]{rotating}

\usepackage[utf8]{inputenc}
\usepackage{fourier} 
\usepackage{makecell}
\usepackage{wrapfig}
\usepackage{tabto}
\usepackage{hyperref}

\usepackage[table]{xcolor}
\hypersetup{
    colorlinks,
    linkcolor={red!50!black},
    citecolor={blue!50!black},
    urlcolor={blue!80!black}
}

\usepackage{tikz}
\usetikzlibrary{tikzmark}

\graphicspath{{figures/}{pictures/}{images/}{./}} 

\begin{document}

\title{Exploring Mid-Air Hand Interaction \\ in Data Visualization} 


\author{Zona Kostic, Catherine Dumas, Sarah Pratt, and Johanna Beyer
\IEEEcompsocitemizethanks{
\IEEEcompsocthanksitem Zona Kostic and Johanna Beyer are with John A. Paulson School of Engineering and Applied Sciences, Harvard University, Cambridge, MA, 02138.
E-mail: \{zonakostic, jbeyer\}@seas.harvard.edu
\IEEEcompsocthanksitem Catherine Dumas and Sara Pratt are with Simmons University School of Library and Information Science, and State University of New York, Albany.
E-mail: \{catherine.dumas2, sarah.pratt\}simmons.edu
}
\thanks{Manuscript received July 17, 2023.}}

\markboth{IEEE Transactions on Visualization and Computer Graphics, November~2023}%
{Shell \MakeLowercase{\textit{et al.}}: A Sample Article Using IEEEtran.cls for IEEE Journals}


\maketitle

\begin{abstract}
Interacting with data visualizations without an instrument or touch surface is typically characterized by the use of mid-air hand gestures. While mid-air expressions can be quite intuitive for interacting with digital content at a distance, they frequently lack precision and necessitate a different way of expressing users' data-related intentions.  
In this work, we aim to identify new designs for mid-air hand gesture manipulations that can facilitate instrument-free, touch-free, and embedded interactions with visualizations, while utilizing the three-dimensional (3D) interaction space that mid-air gestures afford. 
We explore mid-air hand gestures for data visualization by searching for \textit{natural} means to interact with content. We employ three studies---an Elicitation Study, a User Study, and an Expert Study, to provide insight into the users' mental models, explore the design space, and suggest considerations for future mid-air hand gesture design. In addition to forming strong associations with physical manipulations, we discovered that mid-air hand gestures can: promote space-multiplexed interaction, which allows for a greater degree of expression; play a functional role in visual cognition and comprehension; and enhance creativity and engagement. We further highlight the challenges that designers in this field may face to help set the stage for developing effective gestures for a wide range of touchless interactions with visualizations.
\end{abstract} 







\vspace{-3mm}

\section{Introduction}

In recent years, information visualization has become ubiquitous in representing data as a visual means. Visualization is now accessible to a large number of users and in a variety of platforms, such as ambient screens~\cite{stasko_ambient}, wearables~\cite{glanceables}, or immersive environments ~\cite{immersive_survey}. These diverse audiences and spaces necessitate alternative methods of interacting with visualization. Nevertheless, many interaction modalities are frequently neglected in visualization, thereby limiting the ability to discover user intents \cite{interactions_for_infovis}. To investigate engagement in information visualization for a general audience, it is necessary to evaluate different inputs used to perform operations and manipulations. Thus far, visualization has focused on many interaction modalities such as, instrumental interaction \cite{instrumental_interaction_jansen}, proxemics \cite{proxemics}, touch \cite{touchwave}, and speech \cite{audio_vis}, and Lee et al. \cite{beyond} provide a detailed overview of the potential of novel modalities in data visualization. Prior to the emergence of immersive environments and their use for data visualization, however, the significance of mid-air hand gestures was frequently overlooked due to their limited precision and reliability. The advantages and disadvantages of hand gesture interaction are revealed by a comprehensive examination of techniques employed for immersive environments \cite{review_immersive_inter}. Despite technological advancements, there is no standard for ensuring that a set of mid-air hand gestures is appropriate for a domain. Moreover, to provide a consistent user experience, gesture sets should be transferable between domains, and Hosseini et al.'s \cite{towards_a_consensus} study seeks to identify cross-domain consensus of hand gestures in HCI. However, their study does not focus on hand interactions with data visualization. The discoverability of mid-air hand gestures for a set of visualizations and their applicability and usability have not been the primary focus of any visualization or HCI studies. To the best of our knowledge, the appropriateness of this mode of interaction has not been thoroughly researched. 

Compared to conventional interaction modalities, mid-air gestures are mostly utilized in immersive environments, such as interactions with situated visualizations \cite{hybrid}, or interactions with public displays \cite{tangible_public_vis}. In addition to the possibility of expressing physical manipulations while interacting with immersive content, gestures can also be used in a variety of other contexts. For example, the \textit{Gesture Watch} \cite{gesture_watch} enables users to perform mid-air gestures above the watch due to the limited size of its interaction surface. With this, mid-air gestures expand the interaction space, particularly for transitions and changes that do not require high precision (e.g., guides, strokes, and manipulations). In situations involving interactive applications, where touch and speech are inadequate due to occlusion issues (``fat finger''), ambient noise, or the lack of suitable instruments, mid-air hand gestures are a viable touch-free interaction option \cite{smartwatch}, \cite{space_around_smartwatch}, \cite{free_hand_public_display}. This also presents the opportunity for touchless interaction, where certain tasks can be performed more effectively from a distance \cite{freehand_pointing_clicking}, allowing users to move their arms in more flexible and comfortable positions \cite{ARPads}. Besides their versatility, a significant advantage of mid-air gestures is their perceived naturalness and intuitiveness \cite{kinect_leap_myo}, \cite{intuitive_and_natural}.



\begin{figure}[t!]
\centering
\includegraphics[width=\linewidth]{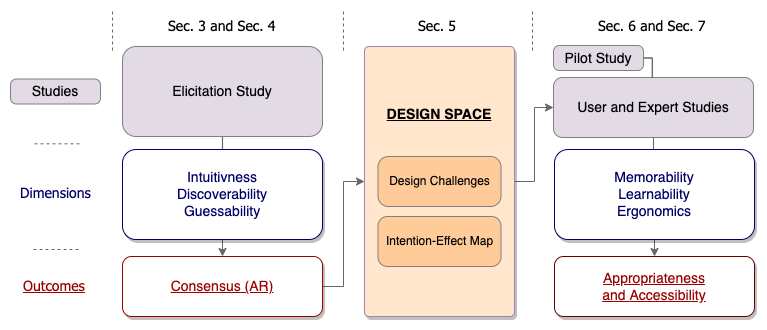}
 \vspace{-7mm}
\caption{Our research methodology. From the elicitation study (left column), we derive and explore the design space for mid-air gestures (middle column). We evaluate the gestures in subsequent user and expert studies (third column). Each column lists \includegraphics[scale=0.5]{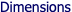} of appropriateness \cite{elic} for gesture selection related to studies, as well as the \includegraphics[scale=0.5]{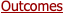} of studies.}
 \vspace{-5mm}
 \label{fig:methodology}
\end{figure}

In this paper, we examine the role of mid-air hand gestures as a touchless interaction modality with data visualization, paying particular attention to previously identified challenges such as recognition, input resolution, and comfort \cite{beyond}, \cite{dynamic_recognition}, \cite{elbow_anchored}. We conducted three studies: an \textbf{Elicitation Study (ES)}, a \textbf{User Study (SS)}, and an \textbf{Expert Study (XS)} (see \hspace{-1mm}~\autoref{fig:methodology}).
Details on our study goals and outcomes, and overall paper structure can be seen in \hspace{-1mm}~\autoref{fig:methodology}.
In the ES, we focus on the principles from Wobbrock et al. \cite{ar_1}, which are based on the guessability technique. We look to reduce the \textit{gulf of execution} \cite{gulfs}, a discrepancy between user intention and user execution. Next, we explore the \textit{design space} of mid-air hand interactions with visualizations. By exploring this design space, we can balance multiple criteria when designing a gesture vocabulary.  In the SS, we assess the \textit{degree of appropriateness} \cite{elic} of the gesture vocabulary designed for a particular scenario. 
In the XS, we use an ``ease-of-use'' measure \cite{review_elicitation_studies} to rate the \textit{ergonomics} of the designed gestures. We follow the intuitive and ergonomic principles from Nielsen et al. \cite{Nielsen} in the SS and XS. We seek to address the following research questions with all of our studies:
\begin{itemize}
\setlength\itemsep{0.1mm}
    \item What is the design space for mid-air hand interactions with data visualizations? (Elicitation Study)
    \item What is the learnability of designed gestures, and how does this affect the memorability for new users? (User Study)
    \item What is the trade-off between memorability and comfort for designed gestures? (Expert Study)
\end{itemize}



The main contribution of the paper attributes to the exploration of a design space of mid-air hand gestures for interacting with data visualizations. Following the exploration, we identify and evaluate a vocabulary of mid-air hand gestures in a user and expert study. Furthermore, we present a comprehensive discussion of the experiences and insights gained during the various stages of design, implementation, and evaluation. Our studies found that gesture suggestions are influenced by the visual encoding used to represent the data. However, the classification of gestures based on physical actions is consistent across visualizations for the same operation. Filtering time, for example, elicits semaphoric dynamic gestures, such as repeated hand flicks, consistently throughout different visual encodings, such as barcharts and scatterplots. In addition, legacy bias and physical actions have a greater impact on elicited gestures than visual representation, which directly affects the learnability and memorability of the gestures for new users. Users prefer gestures with strong physical associations and are able to use parallel inputs and hybrid configurations despite minor ergonomic challenges. To conclude, we provide considerations for mid-air hand interaction design and stress on design tensions and challenges that designers could face in the future. 


\section{Related Work}

Hand gestures serve multiple communication and comprehension functions \cite{hand_and_mind}. An attempt to visualize an object, such as an interface component like a button, is frequently accompanied by gestures that express the object's functionality (pressing or clicking) in addition to outlining its shape (round or square) \cite{colombia_paper}. Hands are frequently referred to as a \textit{natural} mode of interaction because this is how we interact with the physical world and its objects \cite{vuletic}. In popular culture, artists have drawn inspiration from the portrayal of actions using \textit{in-air} or \textit{mid-air} hand gestures for art performances and various science fiction films; this is similar to the manipulations depicted in Steven Spielberg's 2002 film \emph{Minority Report}. 
Hand gesticulation, such as touching the surface, manipulating the object, or communicating via in-air expressions, enables the congruent representation of diverse concepts. It establishes a clear association between the operation and the task. 

Congruent interactions have been explored in a variety of data visualization contexts, particularly to engage with the content on screens  \cite{perin_direct}. The space has been extended to a hand gestural interaction that activates kinetic manipulations and supports multiple inputs in parallel \cite{touchwave}. 
Mid-air hand interaction appears to be preferable in different scenarios to other interaction modalities. For example, Huang et al. \cite{gesture-graph-vr} support the notion that mid-air hand gestures are more effective than a mouse for complex immersive graphs. Hegde et al. \cite{gestAR} posit that mid-air gestures, rather than voice or gaze, are preferred modes of interaction because of the stronger associations users can have through a more natural manipulation and reduced effort when interacting with the environment. 
Even before the invention of gesture recognition with head-mounted displays, complex interactions with virtual tools were facilitated by fingertip positioning and gesticulating in the air \cite{hybrid_feature_tracking}. Mid-air input has proven effective for large amplitude slide gestures in immersive environments and large screens, but also on small screens, as a way to extend the interaction space or to complement other interaction modalities \cite{e_pads}, \cite{air_plus_touch},  \cite{space_around_smartwatch}. 
The location of a gesture can serve as a reminder of its function, and the motion pattern of a gesture does not need to be relearned if it resembles a shape associated with the action \cite{gestures_vs_postures}.

\subsection{Hand Gestures and Data Visualization}

The expression of data-related intentions with data visualizations generally adheres to interactions that resemble communication with physical objects. For example, Thompson et al. \cite{Tangraphe} and Schmidt et al. \cite{multitouch_graph} discuss multi-touch gestures that allow hands to manipulate the visualization through natural movements, enhancing the experience of ``virtual tangibility'' \cite{touchwave}. Rzeszotarski and Kittur \cite{kinetica}, in particular, see the use of ``physics-based affordances'' for direct dialogue with data as an easy method for users to manipulate, analyze, and interact with visualizations. Wall displays are particularly appropriate for hand gesture interfaces, such as gesticulating with tokens \cite{walltokens} or instruments \cite{imaxes-modeless}, but also using mid-air hand inputs \cite{tobias}. Jansen \cite{yvonne_phd} explores various types of data physicalizations that users can interact with directly using their hands; Taher et al. \cite{yvonne_paper} later suggest that this can be accomplished with or without direct touch, such as hovering one's hand over a dynamic data representation. 

Mid-air hand gestures, when combined with touch \cite{touch_or_near_touch}, pen \cite{pen+mid}, voice \cite{voice_gesture_vis}, or virtual pointer \cite{virtual_hand} effectively complement other modalities and play a significant role in a 3D interaction space. Visual feedback stimulates intuitive mid-air gesticulation and serves as an effective compensation for the lack of kinesthetic feedback in mid-air interaction \cite{freehand_pointing_clicking}. 
In addition to translating button presses, augmented reality interfaces require the user to grab elements, pinch to zoom, and rotate knobs \cite{tailor_twist}. Despite the many potential applications for user interactions through mid-air hand gestures, this input method is still only used for immersive visualizations and experiences.

\subsection{Hand Gesture Design}

Designing a gesture system requires consideration of numerous factors in addition to technological constraints. Nielsen et al. \cite{Nielsen} define the core of the human-based approach when developing gestures by recognizing the following characteristics: Gestures should be easy to perform and remember; intuitive; metaphorically and iconically logical towards functionality; and ergonomic, or not physically stressing when used frequently. Understanding users' mental models and cognitive biases aids in recognizing the associations that motivate users to take specific actions \cite{stasko_mental_models}. For example, hand gestures that resemble zooming into content (e.g., ``pinch-to-zoom'') are performed the same way regardless of the size of the content on the screen \cite{pinch_to_zoom_plus}, \cite{mid-air-pan-zoom}, \cite{Tangraphe}; when interacting with remote visualizations, clicking or selecting is typically performed with a ``pinching'' or ``tapping'' in the air  \cite{smartwatch}, \cite{kinect_leap_myo}, \cite{click_mid_air}; \underline{\texttt{zoom}} and \underline{\texttt{pan}} operations are frequently performed together and follow ``grab\&drag'' movements, expressed using various interaction modalities \cite{grab_and_drag}, \cite{aliasing}, \cite{exploring_techiques_boundary}. There is cross-domain consistency for gestures as well as similar themes across the references \cite{towards_a_consensus}. 
Furthermore, mid-air hand gestures provide the possibility to expand mental models and interaction space into parallel, multi-fidelity, and hybrid inputs \cite{3d-spatial-interfaces}, \cite{space-multiplex}, \cite{spacetop}, \cite{hybrid}. 



\textbf{Gesture Studies.} The existing literature on gesture studies and vocabulary design is highlighting iterative, user-focused, factor-centric methodologies \cite{iteratively_designing_vocabulary}. \emph{Elicitation studies} have yielded an impressive amount of knowledge; Villarreal-Narvaez et al. \cite{review_elicitation_studies} offer a comprehensive review of the literature on user agreements, design, participants, and measures in elicitation studies. The goal of a gesture elicitation study is to determine the \textit{first guess} \cite{microsoft}, which categorizes gestures based on specific design criteria, utilizing the aforementioned dimensions of appropriateness \cite{elic} for gesture selection. This is achieved by demonstrating the result of a gesture and then requesting that users perform the action that caused it \cite{touch_gestures}. The criteria of \textit{intuitive or natural} or interacting with little or no instruction is often impacted by \textit{legacy bias}, where gesture proposals are biased by the users' experiences with prior interfaces and technology. Several approaches exist for minimizing the legacy bias, such as \textit{priming}, \textit{partners}, \textit{production} \cite{reducing_legacy}, or \textit{increased production} \cite{increased_production}. Additional techniques, such as priming participants with a \textit{frame}, are a useful design methodology for retaining participants within a particular scenario, thereby removing the reference to previous technology \cite{framed}. However, interaction designers can also use legacy bias to their advantage~\cite{benefiting} when users intentionally suggest movements based on prior knowledge or experiences. Legacy bias can aid users' comfort and satisfaction, especially when designing novel interfaces. According to the Danielescu and Piorkowski's \cite{increased_production} synthesis of prior research, \textit{culturally shared metaphors} may result in higher agreement, improved discoverability, and enhanced learnability. It is crucial to comprehend the equilibrium between leveraging prior experiences and proposing novel designs.

Another approach to eliciting gestures are Wizard of Oz studies, which replicate an automated system by incorporating a human operator, with the objective of minimizing the gulf of execution. In the absence of implementation, this form of research is employed to evaluate expensive concepts and define the scope of the problem. Nevertheless, this particular concept might be impractical for research involving remote participants due to the potential issues of maintaining consistency both across and within participants \cite{pen_and_touch}, as such challenges could be further exacerbated by delays in communication.

In addition to comprehending users' mental models, gesture design should consider technological constraints. For example, technological issues such as accuracy \cite{degrees-of-freedom} and input resolution \cite{beyond} are observed with mid-air interactions. The direction of translation, hand posture, and translation distance are all subject to change \cite{freehand}. Mid-air hand interaction with spatial representations requires a greater degree of freedom of manipulation, which can be effectively carried out with one's hands. Still, it is subject to certain ergonomic restrictions \cite{3d-spatial-interfaces}, \cite{degrees-of-freedom}. To overcome the lack of uniformity and low levels of directness, elicited gestures must be tested with users in various environments, such as varying lighting or screen sizes on personal computers. A well-executed \textit{user study} can aid designers in comprehending the evolution of technique improvements over time and strike a balance between elicited and pre-designed gestures. In addition, it is necessary to evaluate the ergonomics of designed gestures and understand the trade-off between the intuitiveness and accessibility of users' suggestions \cite{ergonomics}. 
In our paper, we conducted an elicitation study, which was subsequently complemented by  user and expert studies. The reason behind several studies is to achieve the best possible integration of design and implementation.


\textbf{Gesture Classifications.}
A large number of gesture taxonomies exist, such as Karam and schraefel \cite{taxonomy_hci}, Wobbrock et al. \cite{touch_gestures}, or Hoffmann et al. \cite{voice-touch-or-mid-air}. More recently, Stachl \cite{tobias} suggested a mid-air gesture taxonomy which adds \textit{static} pose to the previously defined taxonomies. To define the nature of gestures, Aigner et al. \cite{microsoft} analyzed and grouped hand gestures into different \textbf{gesture types}. Their classification combines three gesture taxonomies and focuses on the functional level of gestures, regardless of context. They demonstrate a preference in performing specific physical actions in respect to users' intentions. For example, an intention to select an object is conveyed through pantomimic gestures, such as grabbing, grasping, or tapping the air. Classifying physical actions and gesture types rather than attempting to create a dictionary of gestures for a given set of functions is another trade-off between elicited and designed gestures. In addition, previous research by Pham et al. \cite{aliasing} on augmented reality gestures, Seyed et al. \cite{mde} on gestures for multi-display environments, and Chan et al. \cite{microgestures} on single hand microgestures explore \textbf{gesture themes}. Participants select comparable gestures for fundamentally related tasks, such as various hand expressions of swiping (using one finger, two fingers, or the entire hand) to select a value within a continuous range. Grouping gestures by types and themes focuses on classifying expressions based on physical actions, that refer to similar mental models and facilitate a more accurate interpretation of the elicited suggestion. 
We identify and extract gesture types and themes from our elicitation study and include them in the design space dimensions.

\textbf{Summary.} Although the importance of mid-air hand gesture interaction in data visualization is acknowledged, little progress has been made in comprehending its design and applicability for a set of visualization- and data-centric operations. 
To fill this gap, in our work we conduct an elicitation study to identify and categorize preferred actions for each gesture effect on a data visualization.
To explore the design space of mid-air gestures in data visualization, we build upon prior work to classify physical actions into gesture types~\cite{microsoft}.
Additionally, using the concept of gesture themes, we conceptualize and interpret gestures for more complex tasks, such as actions requiring parameterization.
Based on the findings of our elicitation study, we implement a set of gestures and evaluate them in subsequent user and expert studies.

\section{Elicitation Study (ES)}

This section describes our Elicitation Study (ES) design, implementation, and findings. 
In our ES procedure, we first show the effect of an interaction on a data visualization (known as a \textit{referent}) to each participant and, next, ask them to suggest a gesture intended to produce that effect (examples of ES referents are presented with \hspace{-1mm}~\autoref{fig:ES_questions}). For instance, we ask participants to propose a gesture for selecting individual bars on a bar chart. To control for legacy bias in the ES while providing ample space for users' expression of opinion, we implemented a combination of \textit{production} and \textit{priming} techniques \cite{reducing_legacy}. We were also curious to find out whether or not people were able to draw on previous experiences to learn or discover mid-air gestures \cite{in_the_wild}. For example, we asked users to compare their suggestions with already established mid-air techniques, such as comparing the suggested gesture with the ``pinch'' gesture widely used in immersive spaces (\textit{production}). 
We further asked participants to keep their hands above the keyboard and focus on showing their hands in front of the camera, which is the spatial range where the Leap Motion can sense the gestures (\textit{priming}).

We ran all our studies (ES, SS, XS) remotely via Zoom\footnote{\url{https://zoom.us/.}} with participants' laptops. There are multiple reasons for this strategy. First, \textit{time constraints and user volume}---it would be hard to have 52 participants complete the study in a single lab given public health and social distancing guidelines; Second, \textit{subjects' attention}---we aim to restrict attention to the novelty of the user interface and to eliminate the novelty of surroundings, technology settings or visualizations; And finally, \textit{reducing hand fatigue}---the comfort of one's home or office invites more natural postures (e.g., elbow resting on a table).

\begin{table}
\caption{We map high-level interaction goals \cite{intentions} (intentions), to \textit{how} tasks are executed in terms of methods \cite{brehmer_munzner}. Then, we identify low-level operations that correspond to these methods. For each operation, we create animations of interactions with data visualizations, referred to as referents, to present to the ES participants.}
\scriptsize
\centering
\begin{tabular}{ |p{1.5cm}|p{1.6cm}|p{1.9cm}|p{1.9cm}|  }
 \hline
 \hline
 \thead{Intention} &  \thead{Method} &  \thead{\underline{\texttt{Operation}}} &
 \thead{Referents} \\
  \hline
  
 \makecell{Elaborate} &  \makecell{Annotate} &  \makecell{\underline{\texttt{Hover}}} &
 \makecell{
 \includegraphics[width=10px]{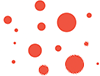}
 \includegraphics[width=10px]{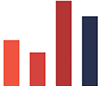}
 }
\\

 \makecell{Select} &  \makecell{Select} &  \makecell{\underline{\texttt{Select}}} &
 \makecell{
 \includegraphics[width=10px]{figures/bubble.png}
 \includegraphics[width=10px]{figures/bar-v.png}
 }
\\

 \makecell{Abstract} & \makecell{Navigate} &  \makecell{\underline{\texttt{Zoom}}} & 
  \makecell{
 \includegraphics[width=10px]{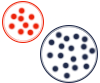}
 \includegraphics[width=10px]{figures/bubble.png}
 \includegraphics[width=10px]{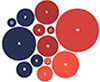}
 }
\\

 \makecell{Explore} & \makecell{Navigate} &  \makecell{\underline{\texttt{Pan}} (drag view)} &   
 \makecell{
 \includegraphics[width=10px]{figures/zoom_clus.png}
 \includegraphics[width=10px]{figures/bubble.png}
 \includegraphics[width=10px]{figures/clusters.png}
 }
\\

 \makecell{Explore} & \makecell{Filter (view)} &  \makecell{\underline{\texttt{Slide}}} & 
   \makecell{
 \includegraphics[width=24px]{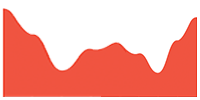}
 }
\\

 \makecell{Explore} & \makecell{Arrange} &  \makecell{\underline{\texttt{Sort}}} & 
   \makecell{
 \includegraphics[width=10px]{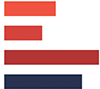}
 \includegraphics[width=10px]{figures/bar-v.png}
 }
\\

 \makecell{Reconfigure} & \makecell{Aggregate} &  \makecell{\underline{\texttt{Reconfigure}}} &   \makecell{
 \includegraphics[width=10px]{figures/clusters.png}
 \includegraphics[width=8px]{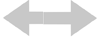}
 \includegraphics[width=22px]{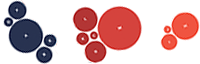}
 }
\\

 \makecell{Reconfigure} & \makecell{Navigate} &  \makecell{\underline{\texttt{Rotate}}} & 
   \makecell{
 \includegraphics[width=10px]{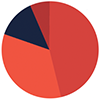}
 \includegraphics[width=10px]{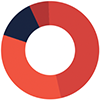}
 }
\\

 \makecell{Encode} & \makecell{Change} &  \makecell{\underline{\texttt{Change}} \\ \underline{\texttt{represent.}}} & 
 \makecell{
  \includegraphics[width=10px]{figures/bar-v.png}
 \includegraphics[width=8px]{figures/arrow_both_direction.png}
 \includegraphics[width=10px]{figures/donut.png}
 }
\\

 \makecell{Connect} & \makecell{Select} &  \makecell{\underline{\texttt{Lasso}}} & 
  \makecell{
 \includegraphics[width=10px]{figures/bubble.png}
 }
\\

 \makecell{Connect} & \makecell{Filter (subset)} &  \makecell{\underline{\texttt{Filter}} \\ \underline{\texttt{category}}} &  
 \makecell{
  \includegraphics[width=10px]{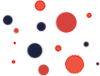}
  \includegraphics[width=5px, height=10px]{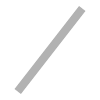}
  \includegraphics[width=10px]{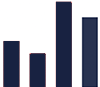}
  \includegraphics[width=10px]{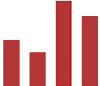}
 }
\\

 \makecell{Filter} & \makecell{Filter (value)} &  \makecell{\underline{\texttt{Filter data}}} &  
 \makecell{
  \includegraphics[width=10px]{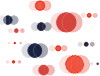}
   \includegraphics[width=5px, height=10px]{figures/dash.png}
  \includegraphics[width=10px]{figures/bar-v.png}
  \includegraphics[width=10px]{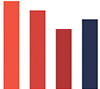}
 }
\\

 \hline
\end{tabular}
\label{table:intention_method_mapping}
\vspace{-5mm}
\end{table}

\subsection{ES Design}

There are many possible interaction \underline{\texttt{operations}} on data visualizations. 
In our ES, we started with high-level user intentions based on Yi et al. \cite{intentions} (such as marking all data points with the same color) and mapped those user intentions to low-level methods and operations as defined by Brehmer and Munzner \cite{brehmer_munzner} (e.g., categorical filtering).

\textbf{\textit{Intention-to-Operation}}. There are numerous taxonomies in data visualization; Dimara and Stasko's \cite{dimara_stasko} work on decision-making provides a comprehensive review of visualization task classification systems. We start with users' data-related intentions based on Yi et al's \cite{intentions} high-level categorization. For instance, if a user is interested in ``exploring'' a set of data points, we can look for hand suggestions they intend to employ. However, their categorization does not establish a clear relationship between intentions, low-level operations, and visualizations. Brehmer \& Munzner’s \cite{brehmer_munzner} typology  bridges the gap between the low-level and high-level tasks focusing on the previous work (including Yi et al.'s classification). To relate intents and interaction techniques, we utilized the \textit{how} portion of the typology (methods that focus on modifying or altering existing data points) that focuses on low-level operations from each group of methods. \hspace{-1mm}~\autoref{table:intention_method_mapping} shows a comprehensive representation of the \textit{Intention-to-Operation} map.

\textbf{\textit{Referents}}. The selection of visual encodings was the first step in determining the referents used in the ES. Each referent is a data visualization with an animation that demonstrates a desired effect, such as selecting a data point. Due to the lack of a clear taxonomy for visual representations, we focused on different visual encodings as defined by Munzner \cite{munzner}, and selected a subset of encodings which are most familiar to a general audience. In addition, we analyzed the choices of representations employed in related studies on interaction design with data visualizations (for example, Saket et al. ~\cite{vis_by_demo}, ~\cite{direct_man_vis}, or Kondo and Collins~\cite{dimpvis}). The final set of referents explore the same operations across different visual representations (e.g., scatterplot \includegraphics[scale=0.4]{figures/bubble.png} and barchart \includegraphics[scale=0.4]{figures/bar-v.png} for \underline{\texttt{select}} operations), as shown in \hspace{-1mm}~\autoref{table:intention_method_mapping}. We wanted to investigate whether different visualizations would elicit different gestures for the same operation (also noted by Willett et al. \cite{elicit-multitouch-selection}). We designed referents without graphical widgets for user interface components to motivate subjects to focus on \textit{embedded} interactions \cite{inter_graphic_encodings}, and address the lack of discoverability in contemporary user interfaces  \cite{nilsen_norman_step_back}.
The referents used in the ES and their corresponding questions are listed in \hspace{-1mm}~\autoref{table:AR_results}.

\vspace{-2mm}

\subsection{ES Participants and Procedure}

The ES included 12 participants. All participants were between 29--60 years old; 8 identified as female; 4 as male; all participants had at least a college degree, 9 had Ph.D. degrees; all were residing in the United States at the time of the study; and none of the participants were HCI or data visualization experts. We recruited participants from the research team's professional network via email.

\begin{figure}
\centering
\includegraphics[width=\linewidth]{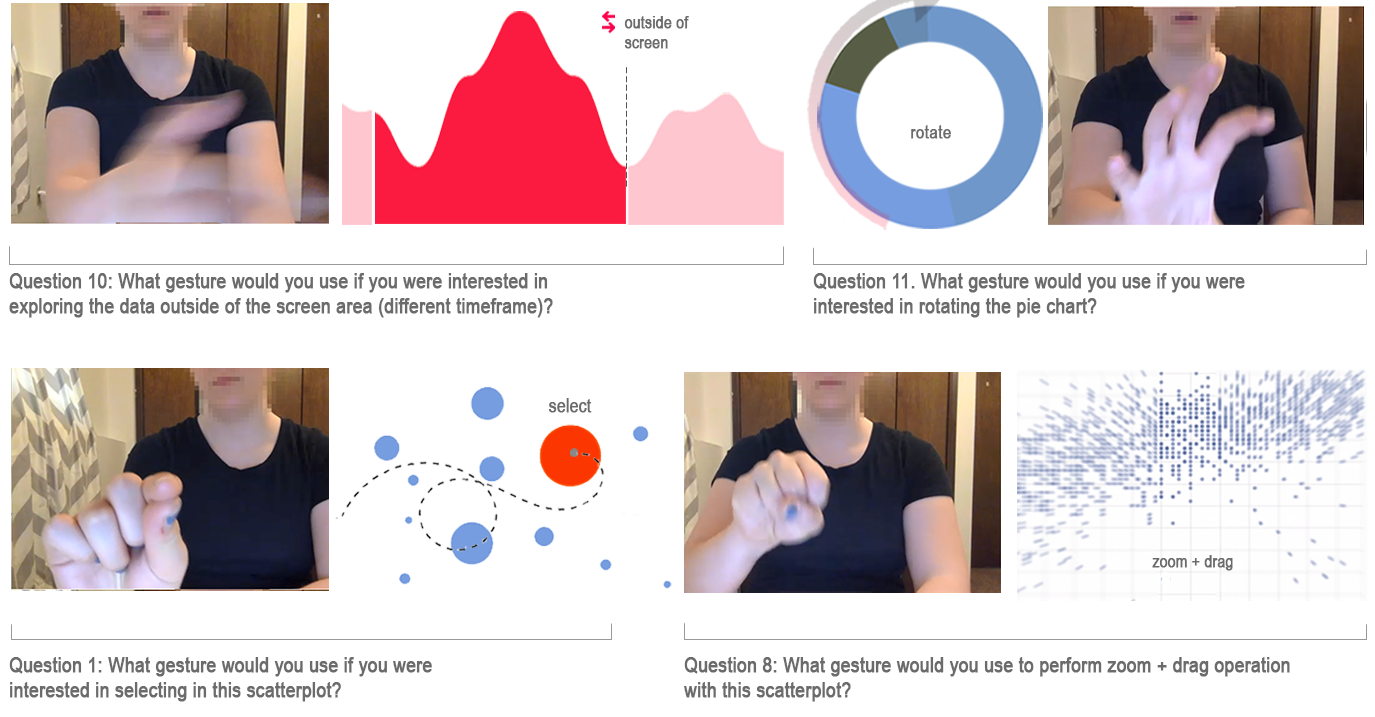}
\caption{Examples of elicitation study questions, referents, and suggestions from a single participant. A complete list of questions can be found in the \tikzmarknode[fill=cyan,fill
opacity=0.1,draw=green!60!black,thick,rounded corners,inner sep=1pt,text
opacity=1]{test}{supplementary materials, Table 1.}}
\label{fig:ES_questions}
\vspace{-5mm}
\end{figure}

We first introduced participants to a series of referents (i.e., visual encodings with an animated operation). Next, we prompted participants to suggest and perform a mid-air gesture to carry out the effect shown. \hspace{-1mm}~\autoref{table:intention_method_mapping} lists all the referents, whereas \hspace{-2mm}~\autoref{fig:ES_questions} illustrates some of the animations in action and the suggestions of a single participant. We presented the referents in increasing difficulty to the participants to encourage them to develop their own mental models and to not overwhelm users. Referents range from simple actions with common visualizations to parameterized actions and transformations not typically encountered by the general public. 
Therefore, all participants completed the tasks in the same order. We asked participants to think aloud and show each gesture in view of their camera and recorded all Zoom sessions for further review. At the conclusion of the sessions, our team shared a post-study survey with participants via email. Participants suggested unimanual and bimanual gestures, and we observed different types of biases.

\subsection{ES Measures}

We use the measure Agreement Rate---$AR(r)$---to calculate gesture agreements across 12 participants. Initially introduced by Wobbrock et al. \cite{ar_1} and later redefined by Vatavu and Wobbrock \cite{ar_2}, $AR(r)$ is a quantitative measure that assesses the congruence between the proposed gestures of several participants. The following formula defines the agreement rate:

$$
\ AR(r) =\frac{\sum_i \sum_{i \neq j} \left[\delta(p_i, p_j) \leq \epsilon \right] }{N(N-1)}\
$$

where $r$ represents the referent, $N$ is the number of participants, $p_i$ and $p_j$ are the proposals of participants $(1 \leq i, j \leq N)$, and $\left[\delta(p_i, p_j) \leq \epsilon \right]$ represents Kronecker's function that evaluates to 1 when the inner expression is true and to 0 when false. $AR(r)$ values range between 0 and 1, from total disagreement to absolute agreement. We used the tolerance $\epsilon$=0 following the previous research on in-air and on-body hand gestures \cite{Wheelchair_users}. To indicate that two gestures are equivalent, handedness, number of fingers, and hand translations must be identical. Vatavu and Wobbrock \cite{ar_2015} delivered a toolkit to assist practitioners to compute agreement rates.

In addition to agreement rates, we categorize gestures by type (\hspace{-1mm}~\autoref{table:gesture_types}) based on the Aigner et al.~\cite{microsoft} classification. The classification of gestures based on their type facilitates the categorization of gestures into themes for individual or multiple operations. 

Three researchers independently analyzed the videos, compared notes, and used a majority vote to classify the gestures into distinct categories. There were no disagreements among researchers. According to Vatavu and Wobbrock \cite{ar_2015}, and later used by Hosseini et al. \cite{towards_a_consensus}, the agreement rate level is to be interpreted as low agreement ($AR(r)$<0.1), medium agreement (0.1<$AR(r)$<0.3), high agreement (0.3<$AR(r)$<0.5), and very high agreement ($AR(r)$>0.5). 

\begin{table}
\caption{The classification of gesture types by Aigner et al.~\cite{microsoft} used in our ES. To identify patterns of gesture types for each referent, gesture suggestions in the ES are classified into the following groups.}
\centering
\begin{tabular}{||l l||} 
 \hline
 Gesture type & Description \\ 
 \hline\hline
 \textit{Pointing} & Indicate objects and directions \\ 
 \textit{Pantomimic} & Imitation, multiple low-level gestures \\ 
 \textit{Semaphoric Static} & Static hand postures (e.g., ``stop'' sign) \\ 
 \textit{Semaphoric Dynamic} & Repeatedly flicking or waving \\ 
 \textit{Semaphoric Strokes} & Single, stroke-like movements \\ 
 \textit{Iconic Static} & Spontaneous static hand postures \\ 
\textit{Iconic Dynamic} & Motions describing paths or shapes \\ 
 \textit{Manipulation} & Motions with subsequent reactions \\ 
 \hline
\end{tabular}
\label{table:gesture_types}
\vspace{-5mm}
\end{table}

\vspace{-2mm}

\section{Elicitation Study Findings}

\begin{figure}[t!]
\centering
\includegraphics[width=\linewidth]{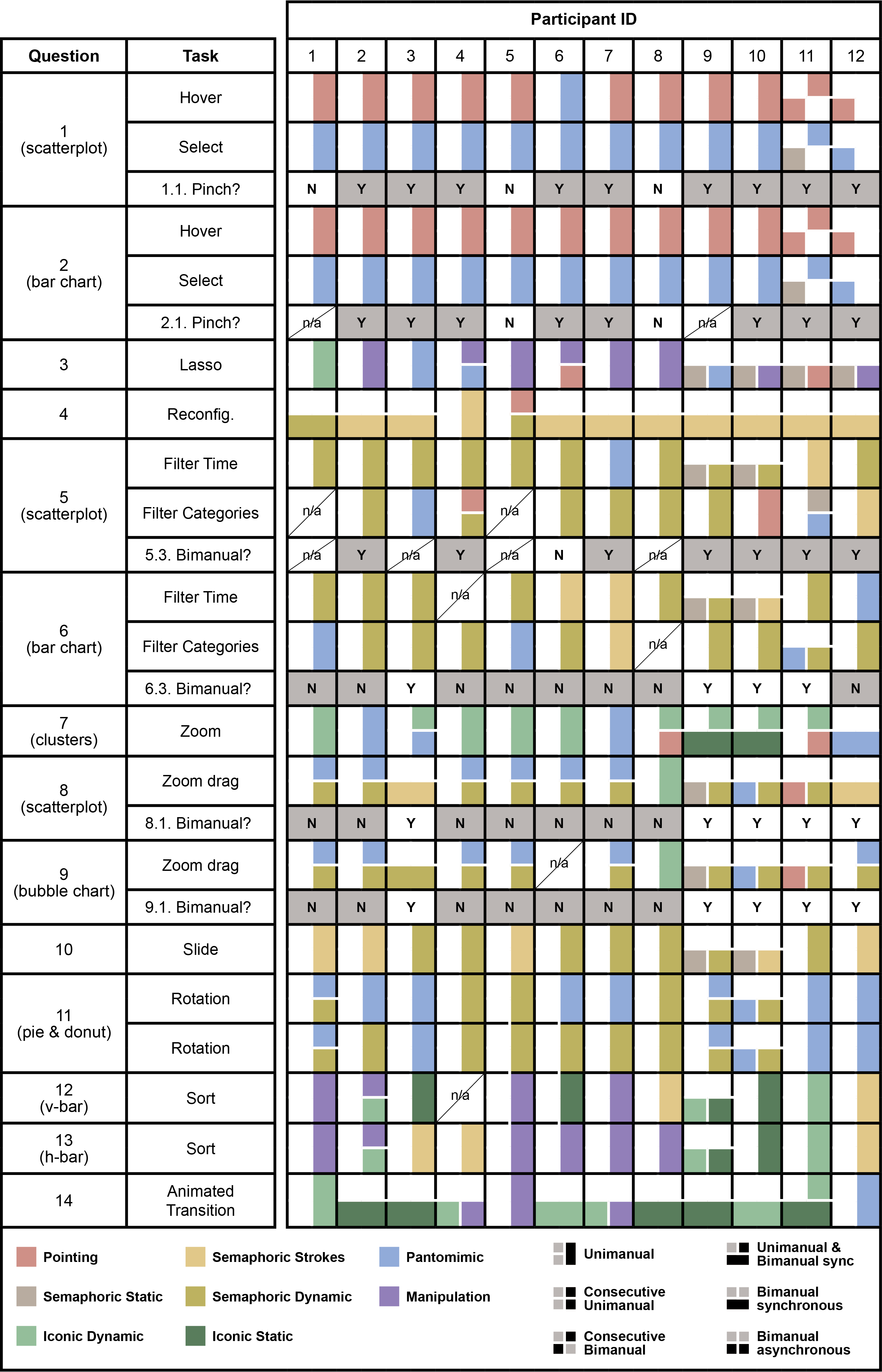}
\caption{\textit{Intention-effect map}. The color represents the gesture type and different patterns show handedness. For more information on gesture suggestions, see \tikzmarknode[fill=cyan,fill
opacity=0.1,draw=green!60!black,thick,rounded corners,inner sep=1pt,text opacity=1]{test}{supplementary materials, Figure 4}}
\label{fig:intention_effect}
\vspace{-3mm}
\end{figure}

As participants frequently described gestures aloud, there were no disagreements regarding the classification of a gesture's effect, type, or theme within our research team. ~\autoref{table:AR_results} shows the calculated agreement rates and~\autoref{fig:intention_effect} shows details on the gesture types participants used. 

Close to fifty percent of displayed referents have a very high degree of agreement ($AR(r)\geq0.5 $). 
For example, participants proposed similar ``index [finger] up'' movements to hover over data points. We found that for the same gesture effect, the translation direction of the proposed gestures varied, and we observed that participants' hands often followed the shape of visual representations (\includegraphics[scale=0.3]{figures/bubble.png}, \includegraphics[scale=0.3]{figures/bar-v.png}). When participants were asked to propose a gesture to reconfigure a bubble chart \includegraphics[scale=0.4]{figures/clusters.png} into clusters \includegraphics[scale=0.4]{figures/clusters_separate_long.png}, nine participants performed a bimanual gesture that mimicked the animated action they observed in the visualization. 
When participants were asked to propose gestures that filter by category in a scatterplot \includegraphics[scale=0.3]{figures/bubble_categories.png}, we observed a heavy influence of the visualization design on the elicited gestures.
We observed, that even for the same operation, the recommended \textbf{gestures vary across different visualization types}. 
For example, when data points appeared layered on top of one another (\emph{z}-axis), participants moved their hands forward and backward to filter data points. 
When working with the pie chart \includegraphics[scale=0.3]{figures/pie.png}, participants' gestures mimicked that of turning a door knob or turning a volume dial, sometimes rotating the whole hand from the wrist and other times using small finger strokes to rotate. Interestingly, gestures between pie and donut chart \includegraphics[scale=0.3]{figures/donut.png} differed, indicating that participants were influenced by the visual encoding as well as the white space in a visualization.

\begin{table}
\caption{Agreement Rates $AR(r)$ for twelve participants ($N$=12), where $P_i$ is the number of different proposals for a referent, and $P$ is the count of the most common identical gesture proposal. The order of the referents is based on the question (Q) asked during the study. The color code displays the following agreement rates: very high \includegraphics[scale=0.05]{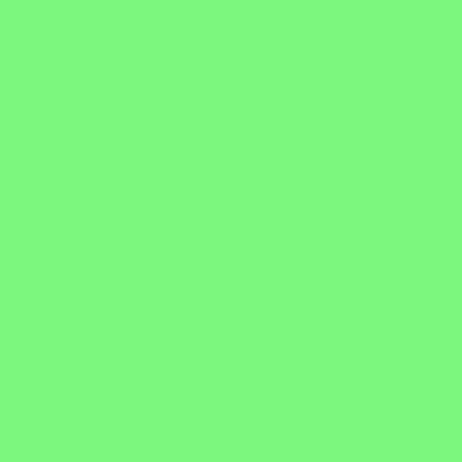}, high \includegraphics[scale=0.05]{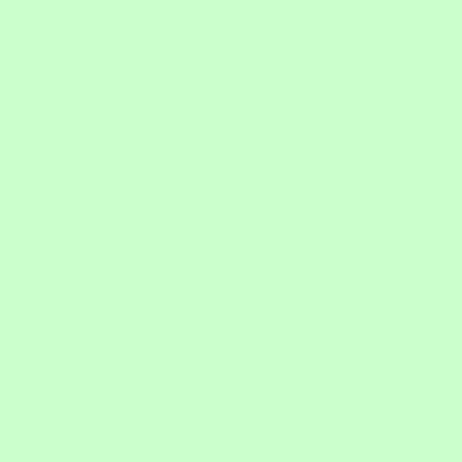}, medium \includegraphics[scale=0.05]{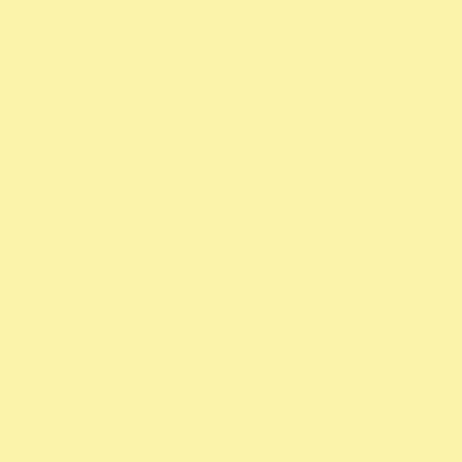}, low \includegraphics[scale=0.05]{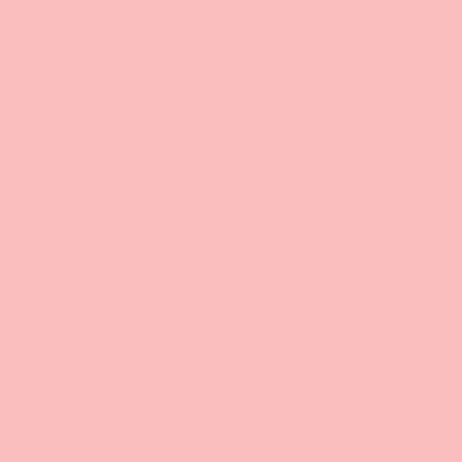}. }
\centering
%
\begin{tabular}{lllll}
 \hline
 & Referents$(r)$ &	$P_i$&	$P$& $AR(r)$ \\ 
 \hline\hline
     Q1 &Hover with scatterplot \includegraphics[scale=0.4]{figures/bubble.png} &3 &8 & \cellcolor{green!40} 0.5\\
      &Select with scatterplot \includegraphics[scale=0.4]{figures/bubble.png}	&3	&6	& \cellcolor{green!20} 0.3\\
     Q2 &Hover with bar chart \includegraphics[scale=0.4]{figures/bar-v.png}	&2	&9	& \cellcolor{green!40} 0.6\\
&Select with bar chart \includegraphics[scale=0.4]{figures/bar-v.png} &3	&6	& \cellcolor{green!20} 0.3\\
    Q3 &Lasso with scatterplot \includegraphics[scale=0.4]{figures/bubble.png}	&4	&7	& \cellcolor{green!20} 0.4\\
    Q4 &Reconfiguration	\includegraphics[scale=0.4]{figures/clusters.png}\includegraphics[scale=0.4]{figures/arrow_both_direction.png}\includegraphics[scale=0.4]{figures/clusters_separate_long.png}			&4	&9	& \cellcolor{green!40} 0.6\\
    Q5.1 &Filter time with scatterplot \includegraphics[scale=0.4]{figures/scatter_move.png}		 &3	&10	& \cellcolor{green!40} 0.7\\
    Q5.2 &Filter categories with scatterplot \includegraphics[scale=0.4]{figures/bubble_categories.png}		 &2 &8 & \cellcolor{green!20} 0.4\\
    Q6.1 &Filter time with barchart
    \includegraphics[scale=0.4]{figures/bar-v.png}\includegraphics[scale=0.4]{figures/arrow_both_direction.png}\includegraphics[scale=0.4]{figures/bar_v_move.png}			&4	&6	& \cellcolor{green!20} 0.3\\
    Q6.2 &Filter categories with barchart \includegraphics[scale=0.4]{figures/bar_blue.png} \includegraphics[scale=0.4]{figures/bar-red.png}	&3	&8	& \cellcolor{green!20} 0.4\\
    Q7 &Zoom into clusters \includegraphics[scale=0.4]{figures/zoom_clus.png}		&6	&4	& \cellcolor{red!10} 0.1\\
    Q8 &Zoom\&drag with scatterplot \includegraphics[scale=0.4]{figures/bubble.png}		&4	&6	& \cellcolor{green!20} 0.3\\
    Q9 &Zoom\&drag with bubble chart \includegraphics[scale=0.4]{figures/clusters.png}		&4	&5	& \cellcolor{yellow!20} 0.2\\
    Q10 &Slide with area chart \includegraphics[scale=0.4]{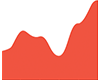}		&2	&7	& \cellcolor{green!40} 0.5\\
    Q11. &Rotate with pie chart \includegraphics[scale=0.35]{figures/pie.png}		&2	&9	& \cellcolor{green!40} 0.6\\
    Q12. &Rotate with donut chart \includegraphics[scale=0.35]{figures/donut.png}		&2	&7	& \cellcolor{green!40} 0.5\\
    Q13. &Sort with vertical barchart \includegraphics[scale=0.4]{figures/bar-v.png}		&7	&3	& \cellcolor{red!10} 0.1\\
    Q14. &Sort with horizontal barchart \includegraphics[scale=0.4]{figures/bar-h.png}		&4	&5	& \cellcolor{yellow!20} 0.2\\
    Q15. &Animated transition \includegraphics[scale=0.4]{figures/bar-v.png}\includegraphics[scale=0.4]{figures/arrow_both_direction.png}\includegraphics[scale=0.35]{figures/donut.png}			&9	&3	& \cellcolor{red!10} 0.1\\
\end{tabular}%
 \vspace{-5mm}
 \label{table:AR_results}
 \end{table}

High agreement rates ($0.3 $<$AR(r) $<$ 0.5$) display a variety of suggestions for the same referent. However, \textbf{the different gestures elicited for the same referent still indicate the same gesture type}. 
For example, participants suggested ``tap'' with index finger only, ``tap'' with the whole hand,  or ``pinch'' (index and thumb) movements when asked what gesture they would use to select a data point on a scatterplot \includegraphics[scale=0.3]{figures/bubble.png}, or a bar chart \includegraphics[scale=0.3]{figures/bar-v.png}; however, all suggestions fall into the same gesture type (pantomimic). To show changes over time in a scatterplot \includegraphics[scale=0.3]{figures/scatter_move.png}, the visualization displayed data points moving left and right (linear), and ten participants proposed a gesture that followed the animation but using different hand symbols. 
Participants either used a single index finger, two fingers, or their whole hand to move left to right, indicating different hand postures, but the same gesture type.
 
When prompted to explore changes over time in a bar chart \includegraphics[scale=0.3]{figures/bar-v.png}\includegraphics[scale=0.3]{figures/arrow_both_direction.png}\includegraphics[scale=0.3]{figures/bar_v_move.png}, participants compared this to scrolling with a mouse wheel or scrollbar widget, \textbf{demonstrating legacy bias}. When prompted to explore zoom\&drag\footnote{Instead of \underline{\texttt{zoom\&pan}}, we are using the \underline{\texttt{zoom\&drag}} operation which is inspired by the actual ``grab\&drag'' gesture \cite{grab_and_drag}.} in a scatterplot \includegraphics[scale=0.4]{figures/bubble.png}, participants followed ``grab\&drag'' movements, but also ``pinch-zoom'' used with touch screens. Despite a great deal of variation in their recommendations, participants suggested the gestures performed with other modalities. This was presented with referents scoring different levels of agreement. For example, when filtering by category in a bar chart \includegraphics[scale=0.3]{figures/bar-red.png}\includegraphics[scale=0.3]{figures/dash.png}\includegraphics[scale=0.3]{figures/bar_blue.png} (\hspace{-1mm}~\autoref{table:AR_results}, Q6.2) participants fixated on interacting with the legend rather than on the \textit{embedded} interaction with the visualization. Moderators prompted participants to consider the legend as static. Still, four participants could not imagine a gesture for filtering in the bar chart when the option of the legend was not available to them. Referents like \texttt{\underline{sort}} with horizontal \includegraphics[scale=0.3]{figures/bar-h.png} and vertical \includegraphics[scale=0.3]{figures/bar-v.png} bar chart as well as animated transition \includegraphics[scale=0.3]{figures/bar-v.png}\includegraphics[scale=0.3]{figures/arrow_both_direction.png}\includegraphics[scale=0.3]{figures/donut.png} (\hspace{-1mm}~\autoref{table:AR_results}, Q12, Q13, and Q14), suggested different variations of drawing and swiping in the air, similar to the gestures demonstrated on touch screens.

The post-study survey at the end of the ES focused on participants' subjective experiences with the novel interaction modality.
Participants reported that they could see themselves interacting using bare hands with various digital resources, \textbf{engaging with data visualizations on home systems}, especially home management systems, as well as entertainment and communication. Opinions regarding mid-air gesture interaction, in general, were somewhat divided between extreme enthusiasm and extreme skepticism, with many participants remarking that it felt artificial and that they would need to become accustomed to the system and its features first. 

\subsection{ES Discussion}

\textit{\textbf{Legacy bias.}} Regardless of the gesture type (e.g., pointing gestures for \texttt{\underline{hover}} operation), gesture theme (e.g., swiping gestures across multiple operations), or agreement rate $AR(r)$, elicited gestures in our study were heavily influenced by the participant's bias. For example, when prompted to suggest a gesture for referents that show \underline{\texttt{hover}} and \underline{\texttt{select}} operations, participants proposed gestures that mimicked the use of a handheld mouse or touch screens; however, participants were also able to adapt to suggestions (e.g., using the ``pinch'' gesture for selection). 
Participants were sometimes distracted by design choices, like the presence of a legend within the various visualizations, and often defaulted to interacting with the legend to manipulate the visualization. Additionally, participants suggested the full-hand pinch gesture for zoom and a swiping motion for any scrolling gesture, which mimics interactions with touch screens. Bias was also present due to natural physical or cultural and social gestures, such as putting one hand up parallel to the screen to signal ``stop.'' Certain referents (e.g., \underline{\texttt{reconfigure}} \includegraphics[width=10px]{figures/clusters.png}
 \includegraphics[width=10px]{figures/arrow_both_direction.png}
 \includegraphics[width=20px]{figures/clusters_separate_long.png}) were not anchored in legacy bias, but still had a high agreement rate which we attribute to the connection of the gesture to a physical action. 
 
\textit{\textbf{Awareness of scale and white space.}} Participants displayed an awareness of the use of space and white space in the visualizations. Participants often changed gestures for the same effect depending on the size of data points, the amount of white space in the visualization, and the level of specificity required by the prompt. This spatial awareness translated to an awareness of the three-dimensional interaction space and bimanual parallel inputs. 

\textit{\textbf{Interface limitations.}} Similarly, participants were influenced by the simulated sequence of the animated visualizations; one participant suggested that since the visualization already had an inherent logic that displayed an action in a certain way, this limited both their imagination to suggest gestures and the functionality of the visualization itself. During the ES, participants demonstrated an understanding that systems are programmed and therefore limited to only responding to or performing certain functions or actions. 
 
\textit{\textbf{Mental models.}} Participants were often concerned with the application itself, what the system would allow them to do, and how it would react and respond. Across all participants, moderators noticed a desire for more context and imposed limitations. Moderators observed that participants often developed their own gesture vocabulary as they moved through the visualizations and task sets. Participants were influenced by their own suggestions from an earlier visualization, either to re-use a gesture to carry out a similar function or not to use a gesture because it had already been exhausted. Participants discussed the issue of disambiguation, desiring a defined set of gestures to control the system and also that gestures are distinct from each other. This was evident in observing participants' self-designed gesture set as they progressed through the study. 

\subsection{ES Limitations}

For a complete comprehension of the ES and its results, we must also discuss its limitations. Overall, participants in the ES were not advised whether or not mid-air hand gestures should be consistent across all required interaction scenarios. This, however, was omitted on purpose because we wanted to determine whether or not users would be inspired to suggest a single gesture for multiple operations or if they have some sort of comprehension of the need for disambiguation. The elicitation study has limitations in the following aspects: 

\textit{\textbf{Recruitment.} }The recruited subjects were professors and graduate students in academia. Although none of them have HCI or data visualization experience, expanding the ES beyond academic audience may be beneficial. 

\textit{\textbf{Visualization and cognition.}} 
We designed visualizations together with a simulated action or animation to elicit gestures. However, participants' creativity might have been limited by the animation shown to them. In order to clearly present operations in the elicitation study, some of the indicators were rendered together with the effect (e.g., highlight data to show \underline{\texttt{hover}} and render a cursor to represent the center of a manipulation by hand gestures). These indicators and feedback have strong associations with traditional WIMP interfaces and thus might explain the high consensus on elicited gestures for those visualizations.

\textit{\textbf{Social aspects.}} Besides technology bias, elicited gestures might be subjected to cultural biases as well. All participants of our elicitation study were residing in the United States. Thus, elicited gestures may have been tied to social or cultural norms (e.g., one hand up for ``stop''); we did not request demographic information regarding cultural associations, heritage, race, or ethnicity from participants. Because of this limitation, the cultural influence of persons who have lived outside of the United States (e.g., arms crossed in ``X'' pattern for ``stop'' in Japan) was not considered or captured by our elicitation study. 

\section{Design Space Exploration}

Based on the results of the elicitation study, we aim to define a \textbf{\textit{gesture vocabulary}} for our user study by exploring the design space of mid-air gestures. In particular, we have identified the following dimensions: \textit{gesture type}, \textit{gesture theme}, \textit{space-multiplexing}, and \textit{visual representation}. 
It is important to mention that these dimensions can overlap and are not meant to be exhaustive. Rather, we aim to provide gesture designers with a common vocabulary to balance different aspects in their gesture designs according to the requirements of applications.

\begin{wrapfigure}{l}{0.10\textwidth}
    \vspace{-4mm}
    \centering
    \includegraphics[width=0.10\textwidth]{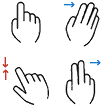}
    \vspace{-8mm} 
\end{wrapfigure}\textit{\textbf{Gesture Types.}} The classification of gestures into gesture types~\cite{microsoft} supports the identification of underlying user intentions and tasks. 
Not only can we draw strong correlations between gesture types and the agreement rates of elicited gestures (for more details, see \tikzmarknode[fill=cyan,fill
opacity=0.1,draw=green!60!black,thick,rounded corners,inner sep=1pt,text
opacity=1]{test}{supplementary materials, Figure 4}), but we can also search for instances in which users perform gestures of the same type while exhibiting posture differences for the same operation. One such instance is \underline{\texttt{reconfigure}}, in which users imitate the separation of bubbles into clusters and demonstrate strong physical associations to the performed action despite variations in translation direction. Regardless of the $AR(r)$ in all tasks, gestures typically fall into the same type category; this categorization is independent of visual representation and serves as a first-level design guideline for gestures suitable for the intended operations. For example, \textit{semaphoric strokes} are hand flicks characterized by single stroke-like movements; their nature, which permits repetitions, suggests interaction with a broader space. Strokes are observed when users intend to bring the data outside of the screen, such as interacting with area charts. On the other hand, \textit{semaphoric dynamic} gestures operate in a more constrained space (e.g., slide). In contrast to dynamic gestures, the actual range of movement in strokes does not convey information regarding the action. \textit{Pantomimic} gestures are used to demonstrate a particular task, such as selecting a single data point, rotating charts, or zooming, regardless of data size. \textit{Manipulation} gestures are used for operations where the effect ``follows'' the gesture, such as sorting operations. Nevertheless, when the hand gesture is aligned with the effect (e.g., animated transition), participants prefer dynamic gestures that are typically used to describe the shape or the transition. \textit{Static} gestures, both \textit{semaphoric} or \textit{iconic}, are observed only when combined with other dynamic gestures across different referents, and are used as a way to manifest the action ``trigger'' (e.g., \underline{\texttt{lasso}}). Participants understood that there must be an initial action (similar to a long click on the mouse to initiate ``drawing'' in the air) and suggested several ways of the same type to imitate this action. 

\begin{wrapfigure}{l}{0.10\textwidth}
\vspace{-4mm} 
    \centering
  \includegraphics[width=0.10\textwidth]{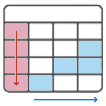}
     \vspace{-8mm} 
\end{wrapfigure}\textit{\textbf{Gesture Themes.}} By mapping gestures to a user's conceptual model, gesture themes assist in identifying commonalities between gestures on a higher, conceptual level.
We discovered two types of gesture themes, \textit{vertical} and \textit{horizontal}. Vertical gesture themes are concepts that present similar gesture suggestions for each referent, and they can be useful for detecting patterns when agreement rates are low. For example, when we ask participants to suggest a gesture for the referent that demonstrates changes over time (filter data), a variety of hand postures are evoked, including ``index [finger] up'', ``pinch'', and ``whole hand''; however, each mimics the use of a slider user interface component with moving the hand ``left-right'' to indicate the desired change. The horizontal concept is exemplified by a consistent behavior across multiple referents. It is useful for identifying identical hand postures and movements across various operations. For example, users ``swipe'' with their hands when they wish to filter time, filter view, or rearrange data. Movements such as ``swipe'' are spatially unrestricted and comprise a series of strokes to bring the data outside the screen. In addition, participants have the same general idea (concept or theme) regardless of the visual representation, which can aid in identifying gesture design patterns applicable to various scenarios. Examples are illustrated in the \tikzmarknode[fill=cyan,fill
opacity=0.1,draw=green!60!black,thick,rounded corners,inner sep=1pt,text
opacity=1]{test}{supplementary materials, Table 4}.

\begin{wrapfigure}{l}{0.10\textwidth}
   \vspace{-4mm} 
    \centering
\includegraphics[width=0.10\textwidth]{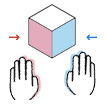}
   \vspace{-9mm} 
\end{wrapfigure}\textit{\textbf{Space-multiplexing.}} With space-multiplexed input, each hand gesture is assigned its own space and channel. Due to legacy bias, the initial suggestions of ES participants contained a few original ideas for parallel inputs (e.g., inputs for multiple operations). Nonetheless, we observed the use of two consecutive gestures in a number of instances, as well as bimanual interactions, indicating that the interaction space of gestures can be expanded to accommodate the simultaneous execution of multiple operations. In the \underline{\texttt{zoom\&pan}} task, for instance, participants opened their hands to perform \underline{\texttt{zoom}} and then made a second gesture to \underline{\texttt{pan}}. Accessing the ``layers'' of data in categorical filtering with scatterplots, involved consecutive spatial movements, which were sometimes performed with both hands accessing multiple categories at the same time. Parallel manipulation reduces the number of distinct postures for a given set of tasks, and the possibility of expanding the interaction space to three dimensions enables hybrid gesture design.

\begin{wrapfigure}{l}{0.10\textwidth}
\vspace{-1mm} 
    \centering
    \includegraphics[width=0.10\textwidth]{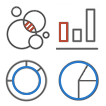}
     \vspace{-8mm} 
\end{wrapfigure}\textit{\textbf{Visual representation.}} The results of the ES indicate that the visual encoding of a referent has a significant impact on the gestures participants chose.
As participants progressed through the tasks, users typically re-used some gestures to perform comparable functions across several tasks. Nevertheless, when the referent visualization was altered, the user responded and modified their gesture. 
In addition, participants concentrated on manipulating only ``visible'' areas of visualizations, and once the visual representation changed, they thought their previous gesture was no longer applicable. 
For rotation tasks, which occurred later in the ES, the visualization itself (donut chart) influenced the participant's gestures as they had the sense that they could not manipulate the white space within the visualization itself. 
In addition to exploring gesture types and themes, physical associations impact gesture suggestions. This idea is consistent with transitions embedded in visual representation. 
In fact, the visual representation triggers these associations, and we can translate these scenarios into appropriate designs. For example, as a method of manipulating orientation, rotation gestures could be applied to any representation provided that it was properly associated with it (e.g., rotate to change orientation, to change direction, or to swap axes). The connection between the physical association and the user's mental model will effectively align with the user's data-related intention.


\begin{figure}[h!]
\centering
\includegraphics[width=\linewidth]{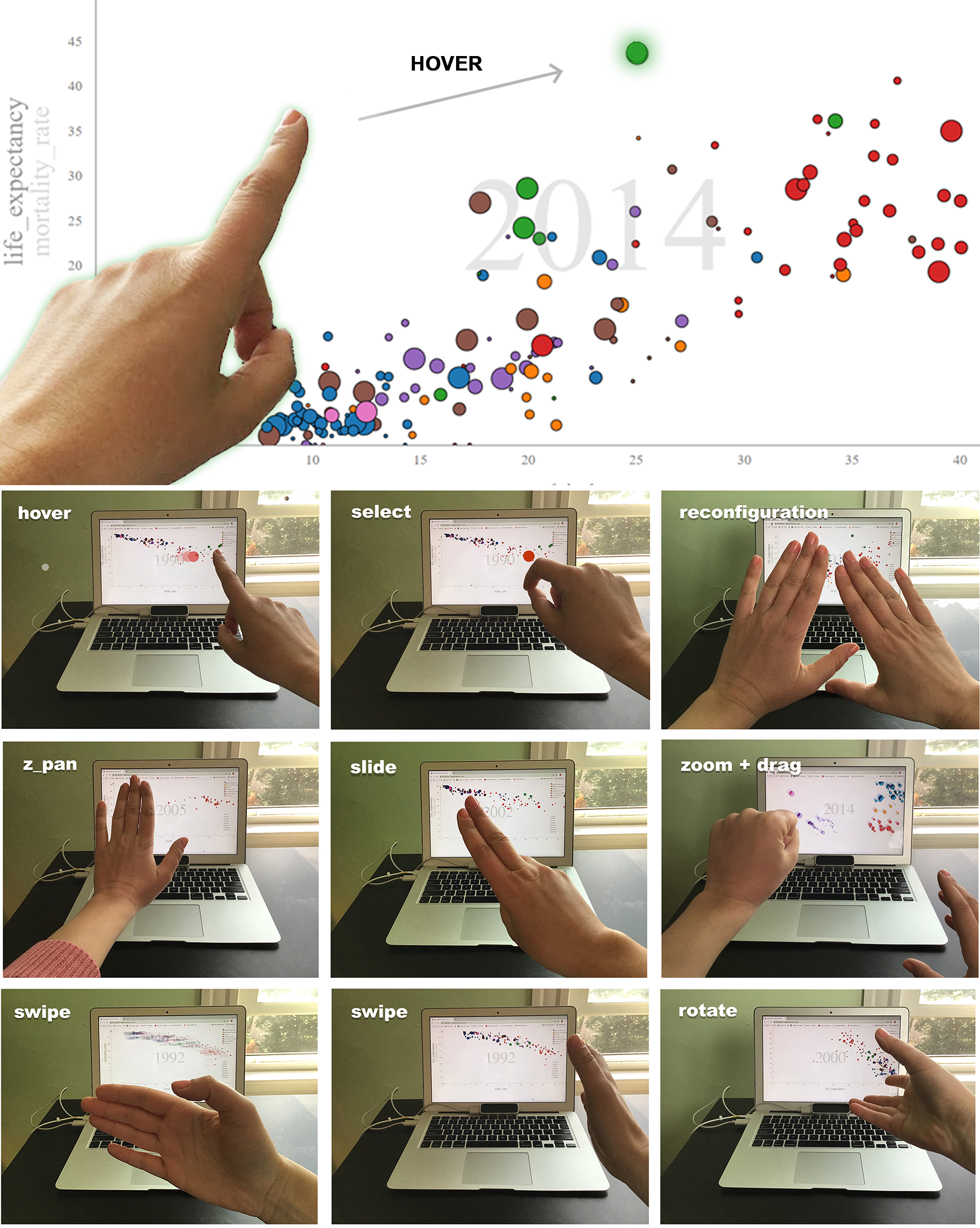}
\caption{The screenshot of the web application running the \textit{Gapminder} visualization with embedded mid–air hand gesture interaction. The presented setup also depicts the equipment used for the user study.}
\label{fig:vocabulary}
\vspace{-5mm}
\end{figure}

\section{Visualization and gesture vocabulary design}
Based on the ES, we implemented different mid-air gestures and evaluated them in a user and expert study. 
To place emphasis on the interaction modality, we chose a visualization and dataset that are well-known to a general audience. Therefore, we used a United Nations dataset, which is also used by the \textit{Gapminder Trendalyzer}\hspace{-1mm} \footnote{\url{https://www.gapminder.org/tools.}}, to populate the visualization. The visualization (referred to as the \textit{Gapminder} visualization throughout the article) encodes four variables of the data in a scatterplot (see ~\autoref{fig:vocabulary}, top). Each data point represents a country, the \emph{x}-axis displays \textit{child birth rate}, the \emph{y}-axis toggles between \textit{life expectancy} and \textit{child mortality rate}, and the size of the data point shows the country's \textit{population size}. Furthermore, countries are grouped by region, which is represented by color. In addition, users can explore changes over time, filter data by category, and swap axes. The visualization design accounts for embedded interaction without any user interface components.

Next, we discuss the vocabulary design for our subsequent studies. We designed gestures with several considerations in mind: $AR(r)$, gesture types and themes, the ability to transition from time-multiplexed to space-multiplexed inputs, and the effect of visual representation. We implemented the following eight gestures: HOVER, SELECT, Z-PAN, SWIPE, SLIDE, ROTATE, RECONFIGURE, ZOOM\&DRAG (\hspace{-1mm}~\autoref{fig:vocabulary}). The majority of gestures (5 out of 8) are designed as elicited, with respect to gesture types and themes. In consideration of technological limitations and disambiguation, we redesigned ZOOM\&DRAG (combining two gestures into one), SLIDE (using two fingers instead of one), and Z-PAN (using the whole hand instead of one finger) to improve the memorability and learnability of the vocabulary, as well as to comprehend the capability for diverse inputs. Utilizing the 3D interaction space maximizes the concept of hand translation direction (e.g., left-right for SLIDE; forward-backward for Z-PAN). To link gestures to physical operations, we looked for the user's intentions expressed in the ES (e.g., for SLIDE, the perception of time is linear, so we use the \emph{y}-axis; for Z-PAN, the \emph{z}-axis is used to access the \emph{layers} of the data participants often perceived with multi-category scatterplots). To avoid confusion, the HOVER gesture employs one-finger manipulations, whereas the SLIDE gesture employs two fingers. ZOOM\&DRAG enables parallel inputs and leaves room for additional operations to be executed. ROTATE is designed as elicited (turning the door nob movement), but it is applied to a new scenario (swap the axes); thus, it adheres to the principle of linking the physical association and the user's initial intent.

\vspace{-2mm}

\section{User and Expert Studies (SS \& XS)}

We conducted a User Study (SS) to explore the intuitiveness of our designed gestures. In an Expert Study (XS), we evaluated the ergonomics of our gestures. We asked participants to analyze hand postures and movements and collected feedback on how to reduce hand fatigue.  
%

\subsection{SS Design and Implementation}

\begin{figure}[t]
\centering
\includegraphics[width=\linewidth]{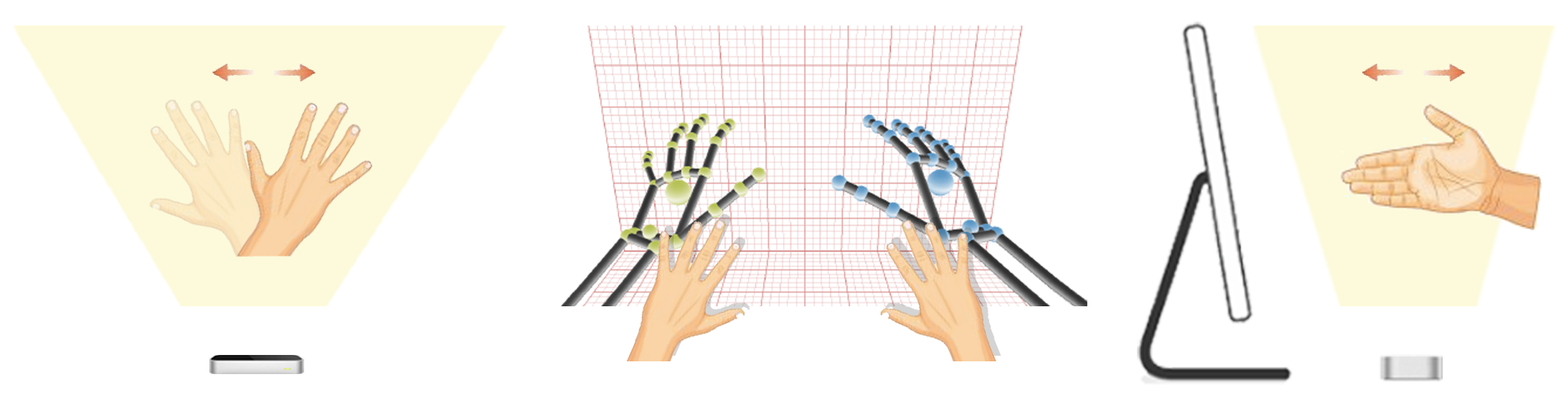}
\vspace{-2mm}
\caption{Tutorial for spatial understanding: Participants practice finding the ``sweet spot'' where the controller can accurately track their hands.}
\vspace{-2mm}
\label{fig:spatial}
\end{figure}

\begin{figure}[t]
\centering
\includegraphics[width=\linewidth]{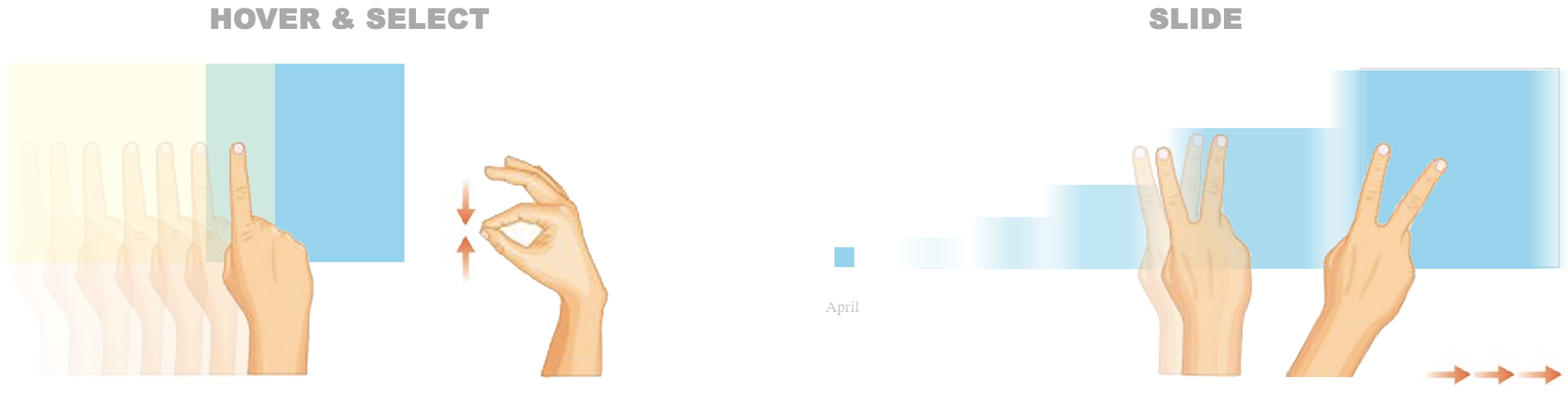}
\vspace{-4mm}
\caption{Tutorial tasks are designed with physical associations in mind and feature distinct visual cues compared to the user study tasks.}
\label{fig:tutorial}
\vspace{-5mm}
\end{figure}

\textbf{\textit{Apparatus.}} We provided Leap Motion Controllers and setup instructions to all participants for the duration of the study (devices were mailed to each participant's home.) 
The setup included a laptop (iMac, Macbook, etc.) and the Leap Motion Controller, connected to the laptop via a USB cable. We developed a web application using \textit{JavaScript} language and Leap Motion SDK\footnote{\url{https://developer.leapmotion.com/.}}. For detecting gestures, we utilized \textit{Frame Objects} and associated attributes to understand the positions of hands and fingers. Visualizations are implemented in D3\footnote{\url{https://d3js.org/.}}.
We ran our study as a web application optimized for a screen resolution of $2,560 \times 1,600$ pixels. The controller's position had to be individually adjusted by each participant. 
The laptop screen sizes varied among the user study participants. 

\textbf{\textit{SS Participants.}} We conducted the user study with 20 participants. All participants were between the ages of 18--49 years; 16 identified as female; 3 as male; 1 identified as non-binary; all participants have at least a college degree; all resided in the United States at the time of the study; and none of the participants were domain experts. We recruited participants from members of the research team's university via email.

\textbf{\textit{Tutorial.}} 
Participants first completed a tutorial 
where we introduced them to the gesture vocabulary and how to position their hands in relation to the sensor (\hspace{-1mm}~\autoref{fig:spatial}). 
The tutorial further introduced simple visual cues (\hspace{-1mm}~\autoref{fig:tutorial}). It encouraged users to feel and learn gestures and make logical connections between a gesture's action and its effect, without introducing any SS-specific tasks. 
After completion of the tutorial, we tested participants on their ability to perform a different variant of a task from the user study. After successful completion, participants moved on to the user study tasks.

\begin{figure}[t!]
\centering
\includegraphics[width=\linewidth]{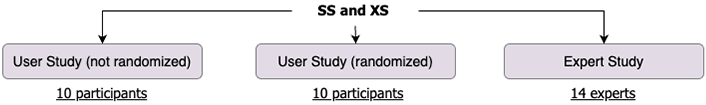}
\caption{Participants and experts in the SS and XS studies.}
\label{fig:us_xs}
\vspace{-5mm}
\end{figure}

\begin{figure*}[tb]
\centering
\includegraphics[width=\linewidth]{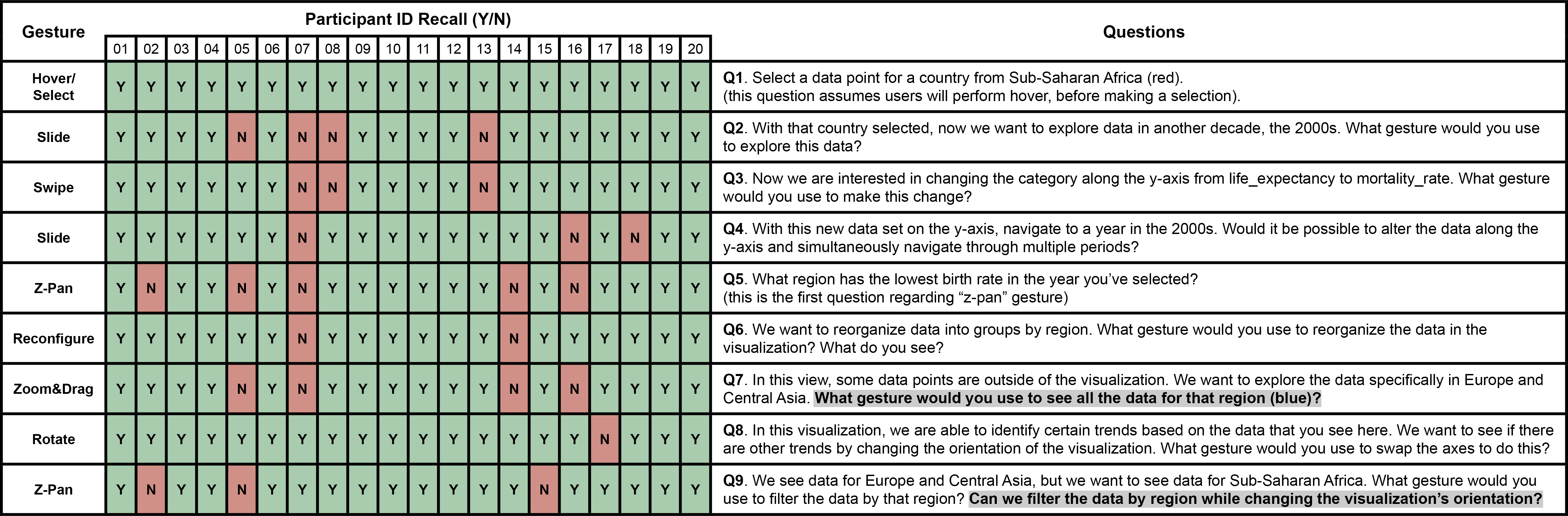}
\caption{The efficacy rates of gesture recall in the user study. Additional questions regarding bimanual manipulation are \colorbox{gray!30}{\textbf{highlighted}}.}
\vspace{-7mm}
\label{fig:us_results}
\end{figure*}

\textbf{\textit{SS Procedure.}} To increase the validity of our study, we randomized the tasks with some users and kept the same order with another group (10 users in each group as shown in \hspace{-1mm}~\autoref{fig:us_xs}). This helped determine whether the order of the tasks in the tutorial had an effect on the recall of the gestures. (We found no difference in recall between the two groups.)
We recorded all user study sessions (including the tutorial and task sets) in Zoom and asked participants to think aloud. We further asked participants to show their hands for each gesture in view of their camera for later review. After completion of the tutorial, we asked participants to complete nine tasks in the user study.
For each task, we showed participants the \textit{Gapminder} visualization and asked them to perform a certain user interaction (e.g., filtering, panning) using the mid-air gesture vocabulary introduced in the tutorial. In addition, we asked participants to execute a number of bimanual interactions with parallel inputs. The complete list of tasks is presented with \hspace{-1mm}~\autoref{fig:us_results}. 
After completing the task set in the user study, we sent a link to a post-study survey to all participants. We asked participants about their experience interacting with the data visualization via gestures (including both Likert scale and open-ended questions) and some demographic questions. 

\textbf{\textit{Pilot study.}} Before running the user study, we conducted a pilot study with six participants who helped us finalize the moderator script and tutorial, and troubleshoot the application. We redesigned the system to improve input reliability and required remote pilot participants to complete the task in their own environment. No changes were made to gestures following the pilot study, and no gender bias was observed among participants. However, pilot participants requested some way to know whether their hands were within the sensor's range or ``sweet spot'', and we tested the system with \emph{virtual hands}. As we progressed with pilot users, we found that participants could easily overcome initial uncertainty regarding new technology and that \emph{virtual hands} occluded visualizations and negatively impacted recall (the participants focused solely on the virtual appearance of their hands on the screen). Therefore, neither the tutorial nor the subsequent studies utilized \emph{virtual hands}. In addition, the pilot study aided moderators in comprehending the need to keep their hands away from the screen and to express the tasks without gesticulation. 

\subsection{XS Design and Implementation}

\textbf{\textit{XS participants.}} We conducted our expert study with 14 experts to evaluate the ergonomics of the eight gestures used in the user study. 
Two co-authors reached out to their domain expert colleagues, including researchers and PhD students.
All XS participants are HCI domain experts (7 postdoctoral fellows, 7 faculty), and eight self-identified as having experience with hand gestures. We solicited participants via private Slack workspaces dedicated to an HCI group of scholars (from the Georgia Institute of Technology, Harvard University, and UC Berkeley).  

\begin{figure}[h]
\centering
\includegraphics[width=\linewidth]{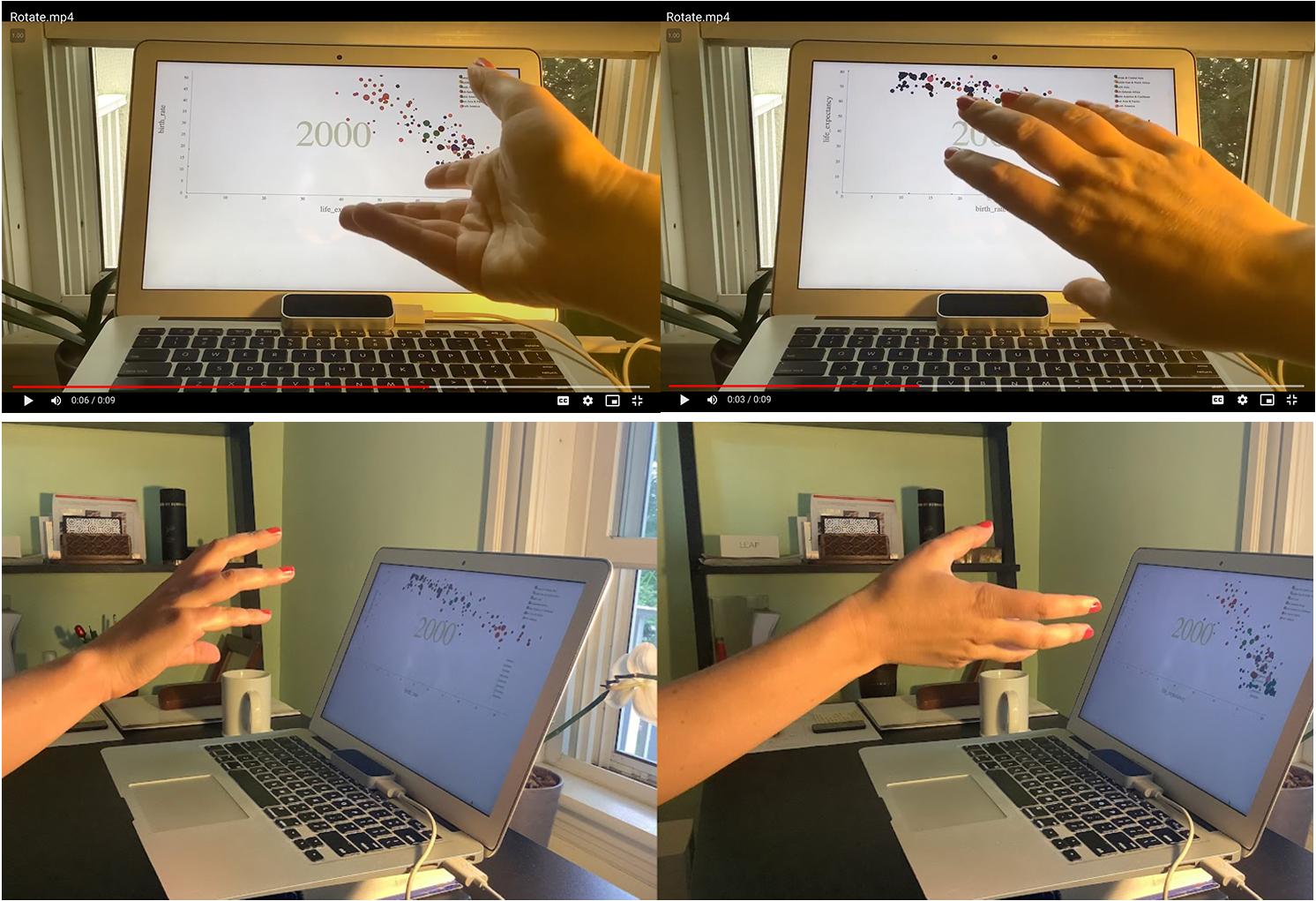}
\caption{In addition to videos, we provided experts with images depicting the same hand posture from different angles.}
\label{fig:xs-side}
\vspace{-5mm}
\end{figure}

\textbf{\textit{XS procedure.}} The expert study consisted of a questionnaire comprised of eight sections, one for each of the eight gestures. We first asked the experts about their general familiarity with gesture and interaction design. Each of the eight sections included a brief video of a user performing gestures on our web application and additional reference images illustrating hand positions from various angles (see ~\autoref{fig:xs-side}). We asked experts to rate the following: discomfort level of hand position relative to their body, muscle strain level resulting from the hand position, and wrist movement. The remaining questions were open-ended and encouraged participant commentary. We list detailed survey questions in the \tikzmarknode[fill=cyan,fill
opacity=0.1,draw=green!60!black,thick,rounded corners,inner sep=1pt,text
opacity=1]{test}{supplementary materials, Table 8 and Table 9}. 

\subsection{SS \& XS Measures}

We evaluate the effectiveness of our designed gestures based on learnability, memorability, and ergonomics (see ~\autoref{table:measures}). 
We analyze \textit{recall} of a novel task variation~\cite{learnability}, \textit{bimanual manipulation} comprehension, and \textit{ergonomics} characteristics evaluated by subject-matter experts. 

After the SS and XS studies, the research team watched the video recordings, calculated the participants' success rates for each task, and observed cognition. Three researchers watched the videos independently and used majority voting for reaching consensus.
The classification of whether a participant correctly recalled a gesture 
was based on: (1) the user took longer than 3 seconds to recall, (2) the user answered with ``I don't know'' or ``not sure'', or (3) the user demonstrated a wrong gesture. 
All research team members agreed on the recall results. 
Additionally, the research team also took note of participants' demonstrating bimanual manipulation and their verbal comments. 


\begin{table}[h!]
\caption{User and Expert studies measures.}
\centering
\begin{tabular}{||l l l||} 
 \hline
 Study & Dimensions & Measures \\ 
 \hline\hline
 SS & Learnability | Memorability & Mapping | Recall \\ 
 XS & Ergonomics Characteristics & Hand Tensions \\ 
\hline
\end{tabular}
\label{table:measures}
\end{table}

\section{User and Expert Studies Findings}
In this section, we report on the findings of SS and XS studies. Detailed results are shown in \tikzmarknode[fill=cyan,fill
opacity=0.1,draw=green!60!black,thick,rounded corners,inner sep=1pt,text
opacity=1]{test}{supplementary materials (Fig. 9, 11, and 12)}. 

\subsection{Learnability and Memorability}

All user study participants completed the tasks in under 15 minutes. The recall success rates, as shown in \hspace{-1mm}~\autoref{fig:us_results}, are depicted with ``Y'' if the participant recalled the gesture correctly on the first attempt and ``N'' if they did not. Almost all gestures have high recall rates, except for the Z-PAN gesture (5 out of 20 users were unable to recall the gesture). The randomization of the task set had no effect on hand gesture recall. Fisher's Exact test shows no significant association between recall count and group association ($p$-value=0.7). On multiple occasions, regardless of whether the question was explicitly asked or not, bimanual manipulation comprehension was observed; when explicitly asked (e.g., Q7 and Q9, \hspace{-1mm}~\autoref{fig:us_results}), all users were able to perform parallel inputs immediately. It is also interesting to observe the results for each user individually; some users had an excellent memory for all gestures, whereas others struggled with the majority of the vocabulary, primarily due to the sensor's input range (e.g., p7 did not recall a gesture in 6 out of 9 tasks).

We observed various demonstrations of visual understanding with different gestures from the vocabulary. This was noted in consideration of both the intuitiveness and logic (metaphorical and/or iconical) of a given gesture. We observed \textbf{increased visual cognition when gestures were related to or mimicked interactions in familiar technology interfaces or the physical world} (e.g., ROTATE). Users commented on an increased understanding of such operations, especially the combination of multiple gestures. For example, parallel operations, such as categorical filtering and axis swapping, make it possible to compare extreme values across multiple groups simultaneously. The presence of embedded interactions in hand inputs allows the user's attention to remain undivided between the visual representation and interface components, thereby facilitating the observation of diverse patterns.

\textbf{Participants felt that interacting with the system using mid-air gestures was easy}, and that the gestures were easy to use. Participants commented on the intuitiveness of the application, saying that they \emph{``thought the gestures were intuitive and did not have a problem using them''}. Other participants stated \emph{``that it feels like a more organic way of interacting with the data''} despite the fact that none of the participants had used this application before. Another participant mentioned that gestures were \emph{``intuitive and matched how you would move data if it were in front of you on a paper''}. Another said that they liked it better than using a mouse and keyboard because \emph{``you can do a lot more with visualizations of data very quickly''}. Participants noted that they enjoyed using the mid-air gestures to navigate the data visualization and said would recommend the system to their friends. Based on the survey responses, moderator observations, and participant comments, all participants found interacting with the gestures and the data visualizations enjoyable and exhibited playfulness.

\subsection{Ergonomics}

Feedback from the XS study indicates that from an accessibility standpoint, all gestures are ergonomically sound (see \hspace{-1mm}~\autoref{fig:xs_ratings}). 
Gesture ergonomic ratings ranged from 1 (low/bad ergonomics) to 5 (high/good ergonomics):
11 (out of 14) experts rated the \textit{posture} of all gestures at four or higher; 
10 experts rated the \textit{muscle strain} of all gestures at four or higher; 
11 experts rated the \textit{wrist movement} of all gestures at four or higher. 
One expert disclosed that they suffer from a chronic illness that can restrict hand use. Their responses and remarks reflect this. In addition to the overall results, participants identified the following characteristics that could improve inclusion and accessibility. Here is how they compare with the efficacy rates of gesture recall:

\begin{wrapfigure}{l}{0.02\textwidth}
    \vspace{-4mm}
    \centering
    \includegraphics[width=18px]{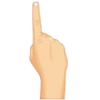}
    \vspace{-9mm} 
\end{wrapfigure} \noindent HOVER. The exceptional memorability of this gesture may conflict with minor ergonomic concerns. The gesture is more challenging for extended periods of time, when using an ergonomic desk setup, or if mobility in the wrist or elbow is limited. Muscle strain could also become an issue if an action requires precision.

\begin{wrapfigure}{l}{0.02\textwidth}
    \vspace{-1mm}
    \centering
    \includegraphics[width=18px]{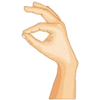}
    \vspace{-9mm} 
\end{wrapfigure} \noindent SELECT. Despite its high learnability, this gesture can be improved ergonomically, as any prolonged hold would be uncomfortable. As a result, users may unintentionally move beyond the system's range, resulting in a loss of accuracy. One participant suggested that users with functional deficits would perhaps benefit from this gesture. 

\begin{wrapfigure}{l}{0.02\textwidth}
    \vspace{-4mm}
    \centering
    \includegraphics[width=18px]{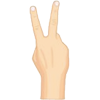}
    \vspace{-9mm} 
\end{wrapfigure} \noindent SLIDE. The gesture is very comfortable and natural, but the tendon strain in the wrist from holding up two fingers is higher than when holding up just one finger.

\begin{wrapfigure}{l}{0.02\textwidth}
    \vspace{-4mm}
    \centering
    \includegraphics[width=18px]{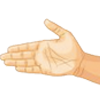}
    \vspace{-9mm} 
\end{wrapfigure} \noindent SWIPE. No recommendations for improvements were suggested as long as the system captures the motion of the hand (movement from wrist) or forearm (from the elbow). Users with carpal tunnel, however, would find this gesture uncomfortable.

\begin{wrapfigure}{l}{0.02\textwidth}
    \vspace{-4mm}
    \centering
    \includegraphics[width=18px]{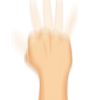}
    \vspace{-9mm} 
\end{wrapfigure} \noindent ZOOM\&DRAG. The complexity of the gesture impacts memorability and muscle strain. Users with functional deficits would perhaps benefit from the ``pinch'' gesture instead. 
If the gesture was combined with a ROTATE gesture, ergonomics could be improved for some users by the system perceiving the rotation of the forearm (rather than the rotation of the wrist) to interact with the visualization.

\begin{wrapfigure}{l}{0.02\textwidth}
    \vspace{-4mm}
    \centering
    \includegraphics[width=18px]{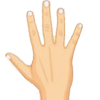}
    \vspace{-9mm} 
\end{wrapfigure} \noindent Z-PAN. This gesture's posture, muscle strain, and wrist movements are rated highly, despite the fact that this gesture has the lowest recall score in the vocabulary. Experts noted that users might experience elbow discomfort if they are too close to the remote display.

\begin{wrapfigure}{l}{0.05\textwidth}
    \vspace{-4mm}
    \centering
    \includegraphics[width=35px]{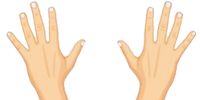}
    \vspace{-9mm} 
\end{wrapfigure} \noindent RECONFIGURE.This gesture was rated as easy to perform, learn, and remember; however, bimanual gestures are limited for users that are one-handed. An alternative gesture would need to be designed as an inclusive and accessible option.

\begin{wrapfigure}{l}{0.02\textwidth}
    \vspace{-4mm}
    \centering
    \includegraphics[width=18px]{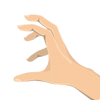}
    \vspace{-9mm} 
\end{wrapfigure} \noindent ROTATE. The gesture is simple to learn and easy to remember, with a high recall rate, but it presents some ergonomic challenges, particularly regarding posture and wrist movement. Users with functional deficits might benefit from a ``pinch'' gesture instead of a ``grab''. One participant commented that the gesture feels ``very normal''. 

\begin{figure}[t]
\centering
\includegraphics[width=\linewidth]{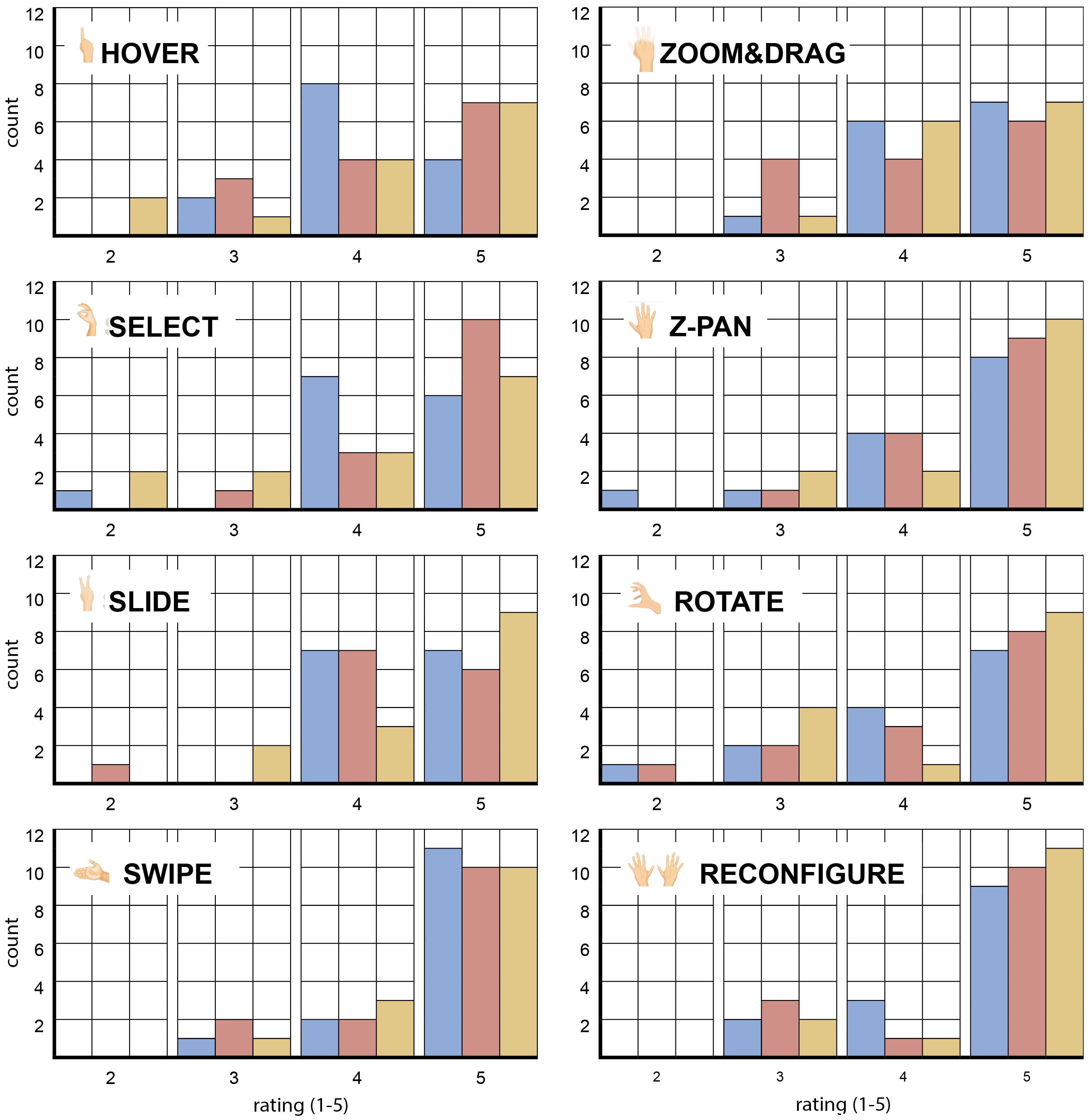}
\caption{The average ergonomics ratings in XS for the following attributes: Posture \includegraphics[scale=0.4]{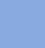}, Muscle Strain \includegraphics[scale=0.4]{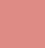}, and Wrist Movement \includegraphics[scale=0.4]{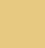}.}
\label{fig:xs_ratings}
\vspace{-6mm}
\end{figure}

\vspace{-1mm}

\subsection{SS \& XS Discussion}

\textbf{\textit{Physical association and legacy bias increase memorability.}} We attribute great recall for HOVER and SELECT to legacy bias due to the existence of similar gestures for other interfaces (e.g., touch screens). Due to the physical gesture mimicking a real-world action, participants also had perfect recall when gestures were mapped to physical actions, such as ROTATE (e.g., turning a door knob). Multiple repetitions of the same gestures throughout the tasks may contribute to the ability to comprehend some bimanual and parallel manipulations, particularly the final one (\hspace{-1mm}~\autoref{fig:us_results}, Q9 that focuses on using ROTATE and Z-PAN at the same time). Nevertheless, given that the questions represent a novel task variation, we are confident that the map between the intention and gesture provides a strong bond and, consequently, strong recall results. Visual cognition was enhanced when gestures related to or mimicked interactions with familiar technology interfaces or the physical world. For example, the SLIDE gesture allows participants to explore changes in data over time. In a traditional WIMP interface, this interaction is often performed using a slider widget and metaphorically maps to the classic interpretation of time being linear. Embedded interaction (directly manipulating the representation) enables the user to focus on the task. At the same time, the direction and speed of the movement can help reveal patterns more quickly as both gesture and data moving manifest the temporal change. This also applies to the RECONFIGURE as an easy switch from the default orientation to a map view. When we explore Z-PAN, the gesture is tied directly to the nature of the visual representation (overlapping data points). Participants observed the ``layers'' of data within the visualization while discovering patterns of specific regions. The ZOOM\&DRAG gesture prompts users to ``grab'' data to move it into view. The operation and outcome mimic the physical act of grabbing an object and bringing it to a space in better view, enabling faster overview-detail comprehension. The ROTATE gesture allows users to change the default orientation of a given visualization. The operation itself (rotating the visualization instead of swapping the axes demonstrated with transition) resembled the physical action of the gesture, which further reinforced the connection between the operation and outcome. 

\textbf{\textit{Subjective feedback indicates high learnability and engagement.}} In a post-study survey, participants were neutral as to whether or not most individuals would be able to learn this method of interaction quickly. However, when prompted to reflect on their own experiences learning the gesture vocabulary, participants suggested it was very easy to learn. Participants felt more engaged with the data, having a sense of controlling the visualization \emph{``physically''}. One participant commented that understanding the 3D nature of the visualization was easier to understand because they were interacting with it through physical gestures. Participants exhibited excitement when interacting with the system and using the novel interaction techniques like Z-PAN, RECONFIGURE, and especially bimanual interaction in both the tutorial and task set. Participants commented on feeling \emph{``powerful''} and some referenced the movie Minority Report. These comments echoed those of the elicitation study, where participants also connected their experience with mid-air gestures to modern cultural access points. 

\textbf{\textit{Technology constraints.}} We asked participants to reflect on what they disliked about using mid-air gestures to interact with the data visualization application. The problems identified are with the system, not the gestures, specifically the frustration the users felt when their hands were not detected by the sensor, primarily during the tutorial phase. We noticed the same inconsistencies between user`s intentions and the lack of systems responses during all of our studies. Most of the problems relate to the issue of \textit{exit error} described by Tuddenham et al. \cite{exit_error} for multi-touch and tangible user interfaces. It refers to a system's difficulty in determining when people switch from non-interacting to action-causing motions. We asked participants to show their hands for each gesture in view of their camera, and we recognize that this prompt may have contributed to participants having their hands outside of the sensor's detection range. Detection and lag time also contributed to an observed decrease in user confidence. Lag time was more prevalent for gestures similar to other gestures in the vocabulary, like HOVER and SLIDE, confirming the need for disambiguation in gesture design. 

\textbf{\textit{Appropriateness and accessibility.}} The discussion surrounding gesture ergonomics is an initial attempt to evaluate the accessibility of designed gesture vocabulary with experts.  In addition, we wanted to investigate how experts' opinions align with those of the elicited suggestions, particularly if any expert ratings contradict the suggested design dimensions. Although not tested in our research, the use of intention-to-operation mapping facilitates the implementation of the same types of gestures for multiple interaction techniques (touch, mid-air, or wearables). This mapping suggests the possibility of achieving uniformity of gestures across multiple modalities for interacting with visualizations. Notably, the positive ratings and high recall rates provided by users further support the viability of the mapping, as most gestures are designed with a specific physical association in mind and align with the predominant gesture types observed during the elicitation study. However, it is worth noting that several recommendations put forth by experts propose the use of the same gesture for different operations, which contradicts the requirement for disambiguation; this is also in line with some of the findings from the elicitation study, particularly when considering user mental models (exhausting the same gesture). In many instances, however, experts suggested that the ergonomics of gestures could be improved by permitting users to hold their hands more loosely to avoid muscle strain. None of the recommendations suggest redesigns that alter the gesture type (for instance, suggestions that lead from dynamic to static gestures) or the need to decompose hybrid inputs into time-multiplexed inputs. 


\vspace{-2mm}

\subsection{SS \& XS Limitations}

Next, we discuss the limitations of our SS and XS studies. 

\textbf{\textit{System setup}}. We ran all the studies remotely, and participants had to install and troubleshoot their setup in advance. Although there are some advantages to this configuration, it is not entirely clear whether these local conditions had a direct relationship with performance and recall. The functionality of the Leap Motion Controller and application are not supported by PCs. Thus our study was limited to participants using Apple machines (iMac, Macbook, etc.).

\textbf{\textit{Recruitment and context.}} Two participants in the user study had prior experience with gesture technology in virtual reality headsets, but none had used hand gestures, and we did not find any differences between them and novices.
Domain experts in the XS study performed gestures after watching recordings. We are unaware if trying the system or watching experts in person would bring different insights and conclusions from these experts. 

\textbf{\textit{Visualization and cognition.}} Even though we utilized the transfer component by using different visual cues for tutorial and user study tasks, we only investigated a scatterplot visualization and did not compare gestures using different visual representations. Furthermore, did not consider the significance or potential effect of color, palette, or intensity design choices in the elicitation or user studies. Since the average length of sessions was 15 minutes, we were unable to investigate the recency effect.

\textbf{\textit{Accessibility.} } The design of the visualization used in the user study revealed an accessibility concern as one participant was colorblind; we did not consider this concern during recruitment. During the user study sessions, the moderator adjusted prompts to accommodate for this. 

\section{Mid-air Hand Gesture Design Considerations}

All three studies (ES, SS, and XS) improved our understanding of mid-air hand gestures used to manipulate data visualizations. These findings, along with the above-mentioned conclusions from prior research, help clarify the relationship between gestures and intended operations and the desire for rules and disambiguation. This section discusses a set of considerations and strategies for resolving future problems that designers may face. 

\textit{\textbf{Appropriateness.}} The gesture vocabulary used in an application must be tailored to the specific purpose of the application. 
When designing a gesture vocabulary for a general audience to interact with casual data visualizations or public displays, designers should rely on legacy bias. We call this a \emph{legacy consensus} as a measure for designing a gesture set that users can easily discover and perform naturally and instantly. For all other applications that necessitate novel gesture designs, designers can rely on the \textit{intention-effect} map, since gesture types are associated with operations regardless of the visualization.

Furthermore, the visual encoding of data, such as the size of data points and the direction of data translation, play crucial roles. According to the ES, scatterplots with numerous small data points lead to single-finger gestures. In contrast, bubble charts with larger data points inspire palm-up gestures for the same operation. This can also be translated to scenarios with different screen sizes. 
Furthermore, animated visualizations present the challenge of translation direction, which is dependent on the data and cannot be predicted in advance. In this case, designers should focus on physical associations to overcome the effect of the translation direction (such as the perception of time being linear). In addition, we recommend that designers encode hand gestures into physical associations, especially for abstract operations. Particularly when the system's reaction occurs after the execution of the gesture, users might form stronger links and even infer novel, unknown parts of the system by performing familiar movements.

\textit{\textbf{Disambiguation.}} One of the strengths of WIMP interfaces is the ability
to disambiguate, whereas gestures are fundamentally less discrete. Users must demonstrate distinct hand movements, which, when enforced on the user's end, have a negative effect on comfort and playfulness. Hand gesture designs must be clearly differentiated from one another, regardless of visual representation, to promote system effectiveness, understanding, and utility. This does not imply that the gesture styles must be distinct, and designers can experiment with various gesture types for different operations. The disambiguation must be present in the direction of translation and distance of designed gestures. These parameters are often observed in individual repetitions in the elicitation study (variability in hand postures directly correlates with memorability). This way, gestures are distinct enough for the user to comprehend the entire vocabulary and for the sensor to differentiate between hand movements more quickly, reducing systems' lag time and enhancing users' confidence.

\textit{\textbf{Input Mapping.}} In practice, the number of operations to perform on a visualization can be quite high, resulting in the development of extensive gesture sets, which directly contradicts the challenge of disambiguation. A 3D interaction space can aid in designing diverse inputs, focusing on the changes in direction translation while keeping the same hand posture. Reducing the number of different postures is also possible by using hybrid techniques. Looking for inputs to be combined into a single technique is challenging. However, the possibility for parallel inputs (combining two distinct operations) may serve as a good starting point. Although hybrid techniques (e.g., \textit{DiveZoom} or \textit{TerraceZoom}\cite{hybrid}) are not the primary focus of our research, we are confident that this space provides an additional opportunity for the development of innovative mid-air interaction designs that can outperform existing modalities.

\textit{\textbf{Accessibility.}} It is important to recognize user group needs in relation to ergonomics. The direction of the gesture itself should hint at the direction of translation or manipulation, but it must be comfortable and discourage fatigue. For desktop settings, gestures should allow users to remain in a static position. Repetitions should be avoided if different muscle groups are required to perform the gesture. If gestures are not ergonomic, it may be necessary to make compromises to promote memorability and reduce fatigue. The gesture vocabulary should maintain spatial consistency that is comfortable for the average user and within reach for both unimanual and bimanual gestures. 

\textit{\textbf{Embodiment.}} The dimension of embodiment describes the degree to which a person perceives technology to be an extension of themselves \cite{beyond}. As a perspective on the relationship between users and systems, embodied interaction \cite{embodiment} implies understanding in both directions: the system's \textit{understanding} of a user`s intent and the user`s comprehension of the system as a ``ready–to–hand'' tool. Every technology has its limitations, which do not need to be explained explicitly to users but must be considered when designing interaction systems. To face challenges with input reliability, designers of mid-air gestures can incorporate embodiment via virtual hands, increasing the user's connection to the system and their confidence. However, the embodiment should correspond visually to the application and not obscure the visualization with opaque objects. In our pilot study, virtual hands distracted users, as they often paid more attention to the hands than the visualization. In immersive representations, particularly in VR, the variety of possibilities is much greater; the appearance of virtual hands does not interfere with the content since the content is in close proximity to the user and is often much bigger than the screens of home users. For desktop applications, incorporating visual responses and highlighting different parts of a visualization with different iconic gestures is a great substitute for virtual hands. For instance, if the system detects a specific gesture, it could highlight the visualization elements that can be interacted with. Visual feedback can subtly alert users when their postures and gestures are becoming ambiguous, thereby guiding their interactions. Ultimately, the user's connection with the system and the system's visual response to a gesture will enhance memorization, cognition, and self-confidence.

\section{Conclusions and Future Work}

This paper explores the design considerations for mid-air hand gestures used to interact with data visualization. We conducted three studies---an Elicitation, User, and Expert Study---to understand user mental models, explore the design space of mid-air gestures, and provide recommendations for future work. We address the following research questions with all of our studies: \textit{What is the design space for mid-air hand interactions with data visualizations? What is the learnability of designed gestures, and how does this affect the memorability for new users? What is the trade-off between memorability and comfort for designed gestures?} 

There is the potential to rely on users' prior experiences and build upon gesture concepts previously explored or learned in other interfaces or interacting with the physical environment. 
We can leverage user knowledge and legacy bias to make novel systems more intuitive. If prior knowledge cannot be effectively utilized, design space dimensions explored in this paper can be used to inform the appropriate design for mid-air hand interaction with visualizations.

We consider our research as a foundation for future work on more specific scenarios. In that respect, future avenues could include collaboratively eliciting gestures by investigating how users perform and interpret a wide variety of operations and strategies through a mass performance study and collective learning. Although we were impressed by the ease of application of the equipment and the way in which Zoom users handled online usability studies, conducting our research in a controlled lab with larger displays may elicit different hand gestures. In order to design gestures that are ergonomic-aware \cite{tailor_twist}, future research should examine the tradeoffs between comfort and precision of mid-air hand interactions. 
Furthermore, we want to investigate gestures for more complex operations and compare them to the findings of this study (e.g., if the gestures differ, are they still the same type?). We are also interested in using \emph{aliasing}~\cite{alias_2} or having multiple \emph{gesture synonyms} for a single operation and seeing how that could affect memorability. Focusing on how designed gestures \emph{feel} in the presence of haptic technology (e.g., \cite{sph}) is an additional dimension that could help mitigate minor accessibility and memorability challenges. Haptic feedback is a complementary modality to mid-air interaction that is used to confirm system response and increase confidence, particularly among visually impaired individuals.




\bstctlcite{IEEEexample:BSTcontrol}
\bibliographystyle{IEEEtranS}
\bibliography{main}

\newpage

\begin{IEEEbiography}[{\vspace*{-5mm}\includegraphics[width=1in,height=1.1in,clip,keepaspectratio]{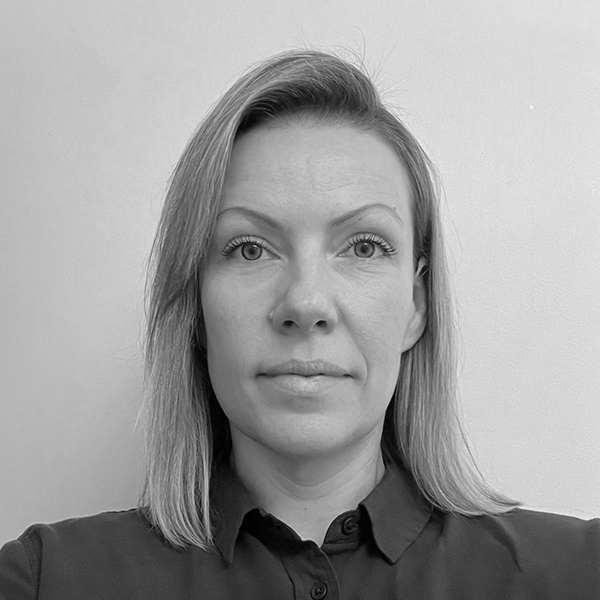}}]{Zona Kostic} is a research fellow and lecturer at Harvard University. Her research lies at the intersection of data visualization, machine learning, and immersive technologies. Zona conducted research on immersive visualization and interfaces with the Visual Computing Group, Harvard SEAS. While pursuing her Ph.D., Zona published six books on computer graphics and design, conducted industry research on human-computer interfaces in virtual reality, and published research on immersive technologies for distance learning. 
\end{IEEEbiography}
\vspace{-10mm}

\begin{IEEEbiography}[{\vspace*{-5mm}\includegraphics[width=1in,height=1.1in,clip,keepaspectratio]{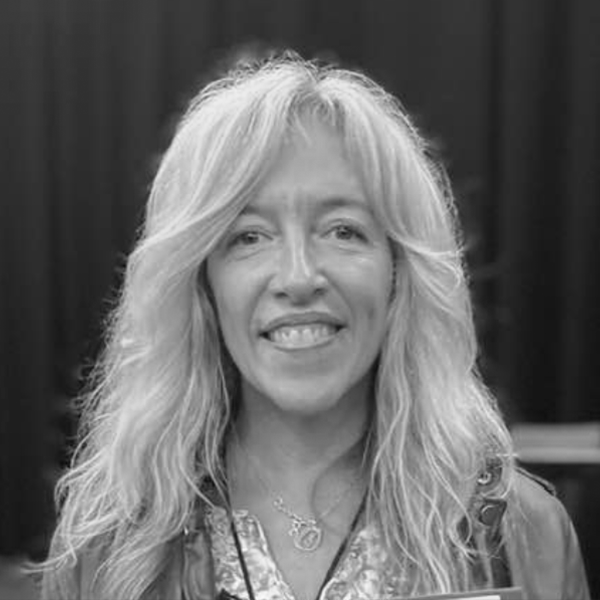}}]{Catherine Dumas} is an Assistant Professor and INF Director at the State University of New York at Albany (Assistant Professor, School of Library and Information at Simmons University at the time of the research). Catherine`s research is inspired by data used for online collective action or digital activism, specifically electronic petitioning. Catherine is currently researching the utility and effectiveness of immersive technologies in education, as well as on the usability, collaboration, and interaction in large virtual reality environments. 
\end{IEEEbiography}
\vspace{-10mm}

\begin{IEEEbiography}[{\vspace*{-5mm}\includegraphics[width=1in,height=1.1in,clip,keepaspectratio]{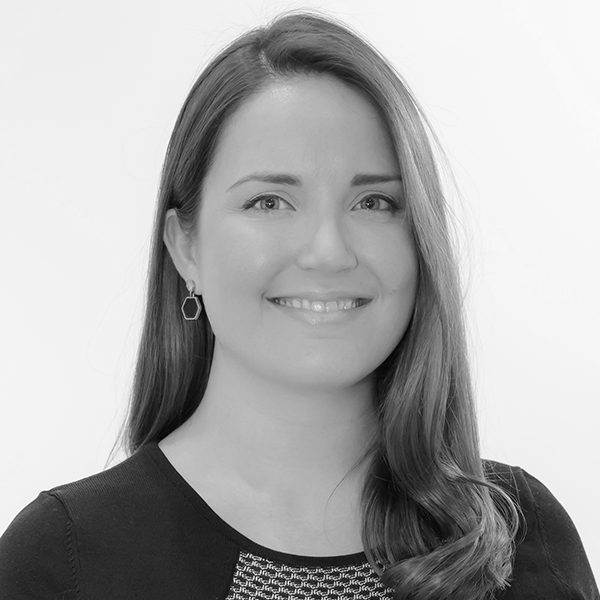}}]{Sarah Pratt} is Associate University Archivist for Community Engagement at Harvard University (Assistant Professor of Practice at Simmons University`s School of Library and Information Science at the time of the research). Sarah is currently serving as co-chair of the Society of American Archivists’ Committee on Research, Data, and Assessment. Sarah received her MSLIS with a concentration in Archival Management from Simmons and has a post-Master’s certificate in user experience research and design. 
\end{IEEEbiography}
\vspace{-10mm}

\begin{IEEEbiography}[{\vspace*{-5mm}\includegraphics[width=1in,height=1.1in,clip,keepaspectratio]{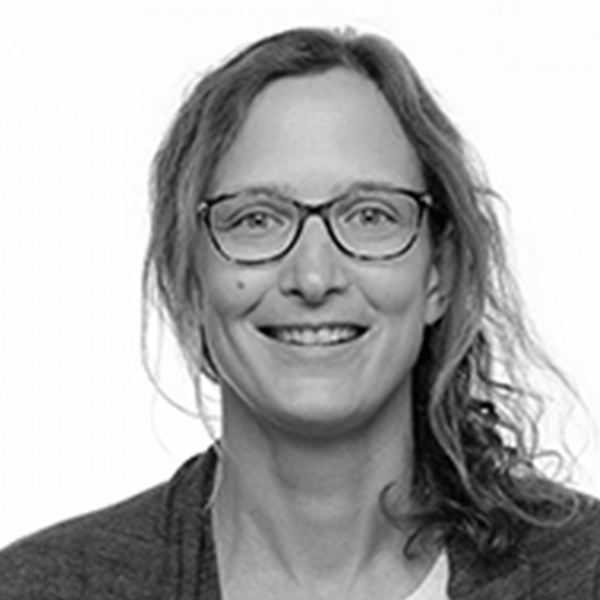}}]{Johanna Beyer} is a research scientist and lecturer at the Visual Computing Lab at Harvard University. Before joining Harvard, she was a postdoctoral fellow at the Geometric Modeling and Scientific Visualization Center at KAUST. She received her Ph.D. in computer science at the University of Technology Vienna, Austria, in 2009. Her research interests include the combination of abstract information visualization with scientific visualization for novel domain-specific applications and immersive visual analytics.
\end{IEEEbiography}

\vfill\clearpage

\end{document}